\PassOptionsToPackage{dvipsnames}{xcolor}
\documentclass[11pt]{article}
\usepackage{jheppub}
\usepackage{bm,bbm}
\usepackage{booktabs,amsmath}
\usepackage{mathtools}
\usepackage{graphics}
\usepackage{braket}
\usepackage{ascmac}
\usepackage{enumitem}
\usepackage{comment}
\usepackage{stackrel}
\usepackage{accents}
\usepackage{simpler-wick}
\usepackage{soul}

\usepackage{framed}
\usepackage{colortbl}
\definecolor{shadecolor}{rgb}{0.90,0.90,0.90}

\usepackage{tikz}
\usetikzlibrary{patterns}
\usetikzlibrary{decorations.markings}
\usetikzlibrary{decorations.pathmorphing}
\usetikzlibrary{decorations.pathreplacing}
\usetikzlibrary{shapes.misc}
\usetikzlibrary{shapes.geometric}
\usetikzlibrary {arrows.meta,bending,positioning}
\tikzset{dot/.style = {circle, fill, minimum size=#1,
              inner sep=0pt, outer sep=0pt}}
\tikzset{cross/.style={cross out, draw, 
         minimum size=2*(#1-\pgflinewidth), 
         inner sep=0pt, outer sep=0pt}}

\edef\restoreparindent{\parindent=\the\parindent\relax}
\usepackage{parskip}
\restoreparindent
\usepackage{ascmac}
\usepackage{mathrsfs}
\usepackage{enumitem}
\newlist{steps}{enumerate}{1}
\setlist[steps, 1]{label = Step \arabic*:}

\def\d{{\rm d}}

\def\i{{\rm i}}

\def\CC{{\cal C}}
\def\CD{{\cal D}}

\def\CI{{\cal I}}

\def\CN{{\cal N}}
\def\CO{{\cal O}}

\def\BR{\mathbb{R}}
\def\BS{\mathbb{S}}

\def\b0{\bm{0}_\perp}

\def\slg{\mathsf{g}}

\def\O{\mathrm{O}}

\usepackage{centernot}
\usepackage{mathtools}
\usepackage{stmaryrd}

\makeatletter
\newcommand{\xMapsto}[2][]{\ext@arrow 0599{\Mapstofill@}{#1}{#2}}
\def\Mapstofill@{\arrowfill@{\Mapstochar\Relbar}\Relbar\Rightarrow}
\makeatother

\usepackage{tikz-3dplot}

\DeclareFontFamily{U}{mathx}{\hyphenchar\font45}
\DeclareFontShape{U}{mathx}{m}{n}{
      <5> <6> <7> <8> <9> <10>
      <10.95> <12> <14.4> <17.28> <20.74> <24.88>
      mathx10
      }{}
\DeclareSymbolFont{mathx}{U}{mathx}{m}{n}
\DeclareFontSubstitution{U}{mathx}{m}{n}
\DeclareMathAccent{\widecheck}{0}{mathx}{"71}
\newcommand{\be}{\begin{equation}}
\newcommand{\ee}{\end{equation}}  
\newcommand{\eps}{\epsilon}
\newcommand{\der}{\partial}
\newcommand{\ie}{{i.e.,}\ }
\newcommand{\no}{\nonumber}

\title{Localized RG flows on composite defects and $\CC$-theorem} 

\author{Dongsheng Ge,}
\author{Tatsuma Nishioka,}
\author{and Soichiro Shimamori}

\affiliation{
Department of Physics, Osaka University,\\
Machikaneyama-Cho 1-1, Toyonaka 560-0043, Japan
}

\preprint{OU-HET-1240}

\abstract{
We consider a composite defect system where a lower-dimensional defect (sub-defect) is embedded to a higher-dimensional one, and examine renormalization group (RG) flows localized on the defect.
A composite defect is constructed in the $(3-\eps)$-dimensional free $\O(N)$ vector model with line and surface interactions by triggering localized RG flows to non-trivial IR fixed points.
Focusing on the case where the symmetry group O$(N)$ is broken to a subgroup $\O(m)\times\O(N-m)$ on the line defect, there is an $\O(N)$ symmetric fixed point for all $N$,  while two additional $\O(N)$ symmetry breaking ones appear for $N\ge 23$. 
We also examine a $\CC$-theorem for localized RG flows along the sub-defect and show that the $\CC$-theorem holds in our model by perturbative calculations.
}

\begin{document}
\maketitle


\section{Introduction}

Quantum field theories (QFTs) host not only point-like particles, but also extended objects such as impurities, boundaries, or interfaces that are ubiquitous in real-world systems \cite{Andrei:2018die}. Those objects in a general sense can all be regarded as defects in the theory.
The presence of a defect introduces a rich variety of new physical phenomena.
They play fundamental roles as charged objects under the generalized global symmetries or generators of those symmetries \cite{Gaiotto:2014kfa}, and have attracted renewed interests in the classifications of the QFT phases \cite{Gaiotto:2017yup}.
Defects cover a subspace of the full spacetime, thus breaking  part of the symmetry of the bulk theory.
The shapes of defects can range from simple planes to more complex structures, thus making it intractable to explore their universal features and dynamics due to the loss of full spacetime symmetry.
While general defects are out of control, topological ones or those preserving a part of spacetime symmetry can be tamed.
Symmetry defects are the examples of the first class while planar defects fall into the second class with the Poincar\'e symmetry preserved only on the defects.
In the latter case, the residual symmetry enhances to the conformal group on the subspace when both the bulk and defect theories are at criticality simultaneously and described by conformal field theories (CFTs).
Such defects are termed conformal defects and CFTs hosting conformal defects are referred to as defect CFTs (DCFTs).

In defect CFTs, a new set of local operators localized on the defects can be introduced. Their presence alters the correlation functions of the bulk local operators,
leading to new critical exponents and scaling behaviors whose kinematical structures are governed by the residual conformal symmetry \cite{Cardy:1984bb,McAvity:1995zd,Kapustin:2005py,Liendo:2012hy,Billo:2016cpy,Gadde:2016fbj,Lauria:2017wav, Lauria:2018klo,Kobayashi:2018okw,Guha:2018snh, Nishioka:2022ook}.
Defect CFTs incorporate the effects of impurities and boundaries on critical systems, making them essential for understanding the renormalization group (RG) flows structures of QFTs with defects around the fixed points,  where conformal symmetries are expected to emerge \cite{Andrei:2018die}.
These RG flows not necessarily occur globally in the bulk, but actually can be localized on the subspace. This is made possible by introducing localized interactions on the defects to deform the theory.
In such cases, the number of the effective degrees of freedom localized on the defect is expected to decrease monotonically under the flow.
This expectation is formalized as the defect $\CC$-theorem, which is proved in specific dimensions and leads to strong constraints on the dynamics of the system \cite{Kobayashi:2018lil, Nishioka:2021uef,Sato:2021eqo, Yuan:2022oeo, Cuomo:2021rkm, Casini:2022bsu, Jensen:2018rxu,  Wang:2021mdq, Casini:2023kyj, Harper:2024aku}.
Meanwhile, defect CFTs provides a window into the dynamics of more exotic setups, such as string theory and holographic dualities (see e.g., \cite{Polchinski:1995mt, Recknagel:1997sb,Alekseev:1998mc, Recknagel:1998ih, Schomerus:1999ug, Bachas:2001vj, DeWolfe:2001pq, Karch:2000ct, Aharony:2003qf, Bak:2003jk, DHoker:2007zhm,Azeyanagi:2007qj,Takayanagi:2011zk,Fujita:2011fp,Nozaki:2012qd,Bachas:2012bj,Jensen:2013lxa,Bachas:2020yxv}).
There have also  been extensive studies on defect CFTs in recent years from various perspectives \cite{Lemos:2017vnx, Liendo:2019jpu,Ghosh:2021ruh,  Yamaguchi:2016pbj, Nishioka:2022odm, Nishioka:2022qmj,Gaiotto:2013nva,Bianchi:2021snj,Gimenez-Grau:2021wiv,Giombi:2021uae,Collier:2021ngi,Soderberg:2021kne,Bissi:2022bgu,SoderbergRousu:2023zyj,Popov:2022nfq,Brax:2023goj}.
 
While models with a single defect (such as boundaries \cite{McAvity:1993ue, McAvity:1995zd,Herzog:2019bom,Herzog:2020lel,Herzog:2022jlx,Herzog:2023dop,Bartlett-Tisdall:2023ghh,Bissi:2018mcq,Prochazka:2019fah,Prochazka:2020vog,DiPietro:2019hqe,Behan:2020nsf,DiPietro:2020fya,Behan:2021tcn,DiPietro:2023gzi,Giombi:2019enr,Cuomo:2021cnb, Harribey:2023xyv,Metlitski:2020cqy,Toldin:2021kun,Shen:2024ixw}, line defects \cite{Cuomo:2021rkm,Billo:2013jda,Soderberg:2017oaa,Lauria:2020emq,Cuomo:2021kfm,Cuomo:2022xgw,Giombi:2022vnz,Gimenez-Grau:2022czc,Aharony:2022ntz,Aharony:2023amq,Dey:2024ilw}, surface ones  \cite{Herzog:2022jqv,Wang:2020xkc,Cuomo:2023qvp,Shachar:2022fqk,Giombi:2023dqs,Raviv-Moshe:2023yvq,Trepanier:2023tvb,Krishnan:2023cff}) have been the subject of most interests in the previous works due to its tractability, 
there are still room for much more intricate systems where two or more defects coexist. This gives rise to our main concern in this work, we are seeking a model for the simplest case where two defects coexist and the  lower-dimensional one is embedded inside the high-dimensional one. This type of defect is called a composite defect. 
The composite defect can find trace in the experimental studies on materials, for instance the surface Kondo effect of the topological insulator \cite{hagiwara2016surface}. Our theoretical studies might lead to  real-world physics, though it is not clear for the moment. Nonetheless, this constitutes another motivation for this work.

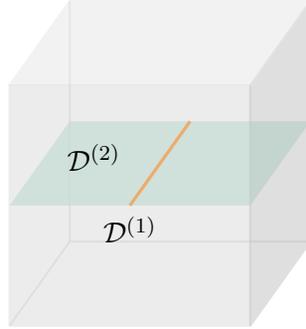
\begin{figure}[t]
  \centering
  \begin{tikzpicture}[thick, scale=1.6]
      \begin{scope}
         \draw[gray!20] (0.5, 2.7) -- (0.5, 0.7);
         \draw[gray!20] (0.5, 0.7) -- (2.5, 0.7);
         \draw[gray!20] (0.5, 0.7) -- (0, 0);
         \draw[draw=none, fill = blue!40!green!30, opacity=0.7] (2,1) -- (2.5, 1.7) -- (0.5, 1.7) -- (0,1) -- (2,1);
         \draw[orange, very thick] (1,1) -- (1.5, 1.7);
         \filldraw[gray!30, opacity = 0.5](0,0) -- (2,0) -- (2, 2) -- (0,2) -- (0,0);
         \filldraw[gray!50, opacity = 0.5](2,0) -- (2.5,0.7) -- (2.5,2.7) -- (2,2) -- (2,0);
         \filldraw[gray!20, opacity = 0.5] (2,2) -- (2.5, 2.7) -- (0.5, 2.7) -- (0,2) -- (2,2);
         \node[below] at (1,1) {$\CD^{(1)}$};
         \node at (0.7, 1.4) {$\CD^{(2)}$};
      \end{scope}
  \end{tikzpicture}
  \caption{The configuration of the composite defect consisting of a line sub-defect $\CD^{(1)}$ (orange) embedded into a surface defect $\CD^{(2)}$ (light green). The gray region is the bulk spacetime where the free $\O(N)$ model lives and localized interactions are turned on only on the subspaces.}
\label{fig:configs}
\end{figure}

From the localized RG flow point of view, once (slightly relevant) localized deformations are turned both on the higher-dimensional defect and the embedded one (the sub-defect), 
a conformal composite defect\footnote{A similar configuration is the wedge CFT, for instance see \cite{Cardy_1983, Antunes:2021qpy, Bissi:2022bgu}.} \cite{Shimamori:2024yms}
arises when the two flows to  non-trivial IR fixed points simultaneously.  We aim to construct a simple  model, where a non-trivial conformal composite defect appears at the Wilson-Fisher fixed point. This brought us to consider the $\O(N)$ vector model. In $d=4-\eps$ dimension, it is known  that a line defect does not have a non-trivial IR fixed point in a free bulk theory \cite{Cuomo:2021kfm}. Embedding such a line defect into a surface one in this dimension does not gain a non-trivial IR fixed point for the line defect either. Instead of turning on the bulk interaction, we consider a lower dimension with $d=3-\eps$. To give a conformal theory at the classical level, the action is required to be
\begin{align}\label{eq:action_intro}
    I
        \equiv 
        \frac{1}{2}\int_{\BR^{d}} \d^{d}x\,\partial_{\mu}\phi^I \partial^{\mu}\phi^I + \frac{h_{\text{0}}}{4!}\int_{\BR^{2}} \d^{2}\hat{y}\, \left( \phi^I \phi^I \right)^2+\frac{g_{0, IJ}}{2}\int_{\BR} \d \tilde{z}\, \phi^{I}\phi^{J} \ ,
\end{align}
where $x$, $\hat{y}$ and $\tilde{z}$ denote the  coordinates for the $d$-dimensional bulk $\BR^{d}$, two-dimensional surface defect $\CD^{(2)}$ and the embedded line defect $\CD^{(1)}$ respectively.\footnote{While we fix the dimensionality of defects throughout this paper, we can  consider the case where the co-dimensions of defects are fixed as in e.g., \cite{Eisenriegler:1988, Prochazka:2019fah}.} The configuration is drawn in figure \ref{fig:configs}. 
In the action \eqref{eq:action_intro}, $\phi^I$ is the O$(N)$ vector field, with index $I$ runs from 1 to $N$;
$h_{0}$ and $g_{0, IJ}$ are the bare couplings for the surface defect and line defect accordingly. We consider the case where the O$(N)$ symmetry is preserved on the surface defect, while on the line it may be broken depending on the form of the coupling $g_{0, IJ}$.

This model allows for perturbative examination both for the surface and line couplings. As either one is turned off, the other one finds non-trivial IR fixed point, proportional to $\eps$ at one-loop order. When both couplings are turned on, the line coupling receives a contribution from the surface coupling and reaches the IR fixed points simultaneously with the surface one. In this sense, a conformal composite defect exists in the model we constructed, even when the bulk theory is free.   This is one of our main results. We have to stress that this is highly non-trivial. If we require that the theory is conformal classically,  dimensional analysis allows only one type of combination in higher even dimension $d=2n -\eps\,(n>2)$, which is to embed an $(n-1)$-dimensional defect into a $(2n-2)$-dimensional one.  In these dimensions,  a conformal composite defect cannot be constructed for a free bulk theory (shown in appendix \ref{app:higerD}). In higher odd dimensions $d=2n-1-\eps\,(n>2)$, two localized classically conformal deformations cannot coexist. This makes $d=3-\eps$ dimensions quite special.

We further analyze the $\O(N)$ structures in the model we constructed, with focus on the $\O(N)$ breaking pattern $\O(N)\to \O(m)\times \O(N-m)$ charaterized by two coupling constants $(g_0,g_0')$ on the line defect. It turns out that conformal composite defects appear not only on the two $\O(N)$ symmetric fixed points, but also on the two $\O(N)$ symmetry breaking ones. Though in the latter case, $N$ is required to be $N\ge 23$ to give a unitary theory. All the IR fixed points share the same value of the   surface coupling $h_*$ at one-loop level,  but differs from each other by  the  set of values of the line couplings $(g_*,g_*')$. To understand the localized RG flows among the all the fixed points quantitatively,  we conjecture a $\CC$-theorem for the sub-defect, or for the submost defect of a system with more coexisting defects. This is a generalization of the defect $\CC$-theorem for a single defect system and serves as the other main result of this work.

The rest of this paper is structured as follows.
In section \ref{sec:single_scalar}, as a warm-up, we consider the RG fixed points and the flows between them in a single scalar field with a composite defect (i.e., we set $N=1$ in the action \eqref{eq:action_intro}). 
In section \ref{sec:ONVecMod}, we analyze to the  $\O(N)$ vector model we have constructed with composite defects, where the O$(N)$ symmetry can be broken on the sub-defect.
In section \ref{sec:sub-defect_C_theorem}, we propose the sub-defect $\CC$-theorem  and check the validity of our conjecture in the model we are studying.
Section \ref{sec:conclusion_discussion} summarizes the main results of this paper and discusses future directions.
In appendix \ref{app:UseFor}, some useful integral formulas needed for the loop calculations are listed.
In appendix \ref{app:higerD}, we show that there are no non-trivial conformal fixed points with composite defects for $d=2n-\epsilon$. In appendix \ref{app:computaton_anomalous_dim}, we give the details on calculating  the anomalous dimensions of the low-dimensional operators.

\section{A single scalar with a composite defect}\label{sec:single_scalar}
In this section, as a warm-up, we consider the model of a single scalar in $d=3-\eps$ dimensions with both a surface defect and a line defect turned on,
\begin{align}\label{eq:actiononeboson}
    I 
        = 
        \frac{1}{2}\int \d^dx\,   (\der \phi)^2   + \frac{h_0}{4!} \int_{\mathbb{R}^2} \d^2 \hat y\, \phi^4 + \frac{g_0}{2} \int_{\mathbb{R}} \d \tilde z\, \phi^2\,,
\end{align}
where $x^{\mu} = (x_\perp, x_\parallel,x_L)$, with $x_\perp$ being the orthogonal direction to the surface defect, $x_\parallel$ being parallel  to the surface but orthogonal to the line, and $x_{L}$ being parallel to the line defect. The 
hatted coordinates cover the surface  defect $\hat y = (x_\perp=0, x_\parallel,x_L)$  and the tilde coordinates cover the line defect $\tilde z = (x_\perp=0,  x_\parallel=0, x_L)$. In this way, the line defect is embedded inside the surface defect. We work in the minimal subtraction scheme; the dimensional regularization allows us to write the bare couplings in terms of the renormalized ones as
\begin{align}
    h_0 &=  M^{2\eps} \left( h + \frac{\delta h}{\eps} +  \frac{\delta_2 h}{\eps^2}+ \dots \right)\,,\\
    g_0 &=  M^{\eps} \left( g + \frac{\delta g}{\eps} +  \frac{\delta_2 g}{\eps^2}+ \dots \right)\,.
\end{align}
To obtain the beta functions of the two couplings, it is enough to consider two types of bulk composite fields, $\phi^2(x)$ and the composite one $\phi^4(x)$. The bare fields and the renormalized ones are related as
\be
    \phi^2 (x) = Z_{\phi^2}\, [\phi^2] (x)\,, \qquad \phi^4 (x) = Z_{\phi^4}\,  [\phi^4](x)\,,
\ee
with $Z_{\phi^2}$ and $Z_{\phi^4}$ being the wave-function renormalizations, however, for free bulk theory $Z_{\phi^2} = Z_{\phi^4} =1$. The physical condition requires that the correlation of the renormalized fields to give a finite answer as $\eps \to 0$, in the case of one-point functions, that is
\be
    \langle\, [\phi^2](x)\, \rangle = \text{finite}\,,\qquad \langle\, [\phi^4](x)\, \rangle = \text{finite}\,.
\ee
Note that only the completely connected diagrams contribute.\footnote{More precisely, the correlation function in presence of the defects is defined as 
\be
    \langle\, \phi^2(x)\, \rangle \equiv \frac{\langle\, \phi(x) D_L D_S\, \rangle }{\langle\, D_L D_S \,\rangle}\,,\qquad
     \langle\, \phi^4(x)\, \rangle  \equiv \frac{\langle\, \phi^2(x) D_L D_S \,\rangle }{\langle\, D_L D_S\, \rangle}\,,
\ee
where $D_{L,S} = e^{-I_{L,S}}$ with $I_{L,S}$ being the action of the line/surface deformation.  In this way, it is clear that  the partially connected diagrams are excluded. 
}
As we shall see, the renormalized couplings at the non-trivial fixed points are of order $O(\eps)$. This renders a perturbative way of calculating the one-point functions.


\subsection{Diagrams for the $\eps$-expansion}
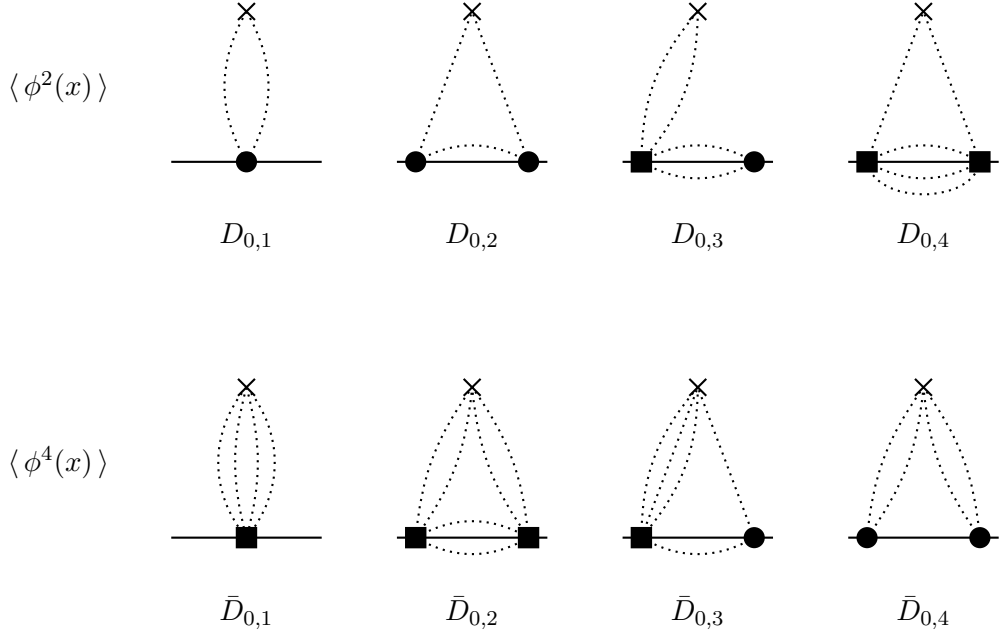
\begin{figure}[t]
  \centering
  \begin{tikzpicture}[thick]
      \node at (-0.5, 1) {$\langle\,\phi^2(x)\,\rangle$};
      \begin{scope}[xshift=2cm]
          \node[cross=4pt] at (0, 2) {};
          \draw[dotted, bend left] (0, 2) to (0, 0);
          \draw[dotted, bend right] (0, 2) to (0, 0);
          \draw[-] (-1, 0) -- (1, 0);
          \node[dot=8pt] at (0, 0) {};
          \node at (0,-1) {$D_{0,1}$};
      \end{scope}

      \begin{scope}[xshift=5cm]
          \node[cross=4pt] at (0, 2) {};
          \draw[-] (-1, 0) -- (1, 0);
          \node[dot=8pt] at (-0.75, 0) {};
          \node[dot=8pt] at (0.75, 0) {};
          \draw[dotted, bend left] (-0.75, 0) to (0.75, 0);
          \draw[dotted] (-0.75, 0) to (0, 2);
          \draw[dotted] (0.75, 0) to (0, 2);
          \node at (0,-1) {$D_{0,2}$};
      \end{scope}

      \begin{scope}[xshift=8cm]
          \node[cross=4pt] at (0, 2) {};
          \draw[-] (-1, 0) -- (1, 0);
          \node[rectangle, draw, fill=black] at (-0.75, 0) {};
          \node[dot=8pt] at (0.75, 0) {};
          \draw[dotted, bend left] (-0.75, 0) to (0.75, 0);
          \draw[dotted, bend right] (-0.75, 0) to (0.75, 0);
          \draw[dotted, bend left=20] (-0.75, 0) to (0, 2);
          \draw[dotted, bend right=20] (-0.75, 0) to (0, 2);
          \node at (0,-1) {$D_{ 1,1}$};
      \end{scope}

      \begin{scope}[xshift=11cm]
          \node[cross=4pt] at (0, 2) {};
          \draw[-] (-1, 0) -- (1, 0);
          \node[rectangle, draw, fill=black] at (-0.75, 0) {};
          \node[rectangle, draw, fill=black] at (0.75, 0) {};
          \draw[dotted, bend left] (-0.75, 0) to (0.75, 0);
          \draw[dotted, bend right] (-0.75, 0) to (0.75, 0);
          \draw[dotted, bend right=80] (-0.75, 0) to (0.75, 0);
          \draw[dotted] (-0.75, 0) to (0, 2);
          \draw[dotted] (0.75, 0) to (0, 2);
          \node at (0,-1) {$D_{2,0}$};
      \end{scope}

      \begin{scope}[yshift=-5cm]
        \node at (-0.5, 1) {$\langle\,\phi^4(x)\,\rangle$};
          \begin{scope}[xshift=2cm]
              \node[cross=4pt] at (0, 2) {};
              \draw[dotted, bend left=15] (0, 2) to (0, 0);
              \draw[dotted, bend right=15] (0, 2) to (0, 0);
              \draw[dotted, bend left=40] (0, 2) to (0, 0);
              \draw[dotted, bend right=40] (0, 2) to (0, 0);
              \draw[-] (-1, 0) -- (1, 0);
              \node[rectangle, draw, fill=black] at (0, 0) {};
              \node at (0,-1) {$\bar D_{1,0}$};
          \end{scope}
    
          \begin{scope}[xshift=5cm]
              \node[cross=4pt] at (0, 2) {};
              \draw[-] (-1, 0) -- (1, 0);
              \node[rectangle, draw, fill=black] at (-0.75, 0) {};
              \node[rectangle, draw, fill=black] at (0.75, 0) {};
              \draw[dotted, bend left] (-0.75, 0) to (0.75, 0);
              \draw[dotted, bend right] (-0.75, 0) to (0.75, 0);
              \draw[dotted, bend left=15] (-0.75, 0) to (0, 2);
              \draw[dotted, bend right=15] (-0.75, 0) to (0, 2);
              \draw[dotted, bend left=15] (0.75, 0) to (0, 2);
              \draw[dotted, bend right=15] (0.75, 0) to (0, 2);
              \node at (0,-1) {$\bar D_{2,0}$};
          \end{scope}
    
          \begin{scope}[xshift=8cm]
              \node[cross=4pt] at (0, 2) {};
              \draw[-] (-1, 0) -- (1, 0);
              \node[rectangle, draw, fill=black] at (-0.75, 0) {};
              \node[dot=8pt] at (0.75, 0) {};
              \draw[dotted, bend right] (-0.75, 0) to (0.75, 0);
              \draw[dotted] (-0.75, 0) to (0, 2);
              \draw[dotted, bend left=20] (-0.75, 0) to (0, 2);
              \draw[dotted, bend right=20] (-0.75, 0) to (0, 2);
              \draw[dotted] (0.75, 0) to (0, 2);
              \node at (0,-1) {$\bar D_{1,1}$};
          \end{scope}
    
          \begin{scope}[xshift=11cm]
              \node[cross=4pt] at (0, 2) {};
              \draw[-] (-1, 0) -- (1, 0);
              \node[dot=8pt] at (-0.75, 0) {};
              \node[dot=8pt] at (0.75, 0) {};
              \draw[dotted, bend left=15] (-0.75, 0) to (0, 2);
              \draw[dotted, bend right=15] (-0.75, 0) to (0, 2);
              \draw[dotted, bend left=15] (0.75, 0) to (0, 2);
              \draw[dotted, bend right=15] (0.75, 0) to (0, 2);
              \node at (0,-1) {$\bar D_{0,2}$};
          \end{scope}
      \end{scope}
  \end{tikzpicture}
  \caption{Diagrams up to the second order in both $h_0$ and $g_0$ contributing $\langle\, \phi^2(x)\, \rangle$ and $\langle\, \phi^4(x) \,\rangle$.
  he dashed line represents the propagator and the cross symbols represent the bulk operator insertions.
  Also, the circle and square represent the line and surface defect interactions, respectively.}
\label{fig:d3o2diags}
\end{figure}

Up to the second order in both $h_0$ and $g_0$, there are four diagrams contributing to the evaluation of one-point function of the quadratic composite operator $ \phi^2(x)  $ and four diagrams contributing to the quartic composite operator $\phi^4(x)$ as well, depicted in figure \ref{fig:d3o2diags}. The basic structure of each diagram, $D_{i,j}$ and $\bar D_{i,j}$  can be written as 
\begin{align}
    D_{i,j} 
        &=
        (-h_0)^i\, (-g_0)^j\, M_{i,j}\, I_{i,j}(x)\,,\\
    \bar D_{i,j} 
        &=
        (-h_0)^i\, (-g_0)^j\, \bar M_{i,j}\, \bar I_{i,j}(x)\,,
\end{align}
where the indices $i$ and $j$ denote the $i$-th order in $h_0$ and the $j$-th order in $g_0$, $M_{i,j}\, (\bar M_{i,j})$ denote the multiplication factor and $I_{i,j}\, (\bar I_{i,j})$ denote the integrals of the propagators over the internal coordinates on the defect surface or line. The multiplication factor $M_{i,j}$ and  $\bar M_{i,j}$ can be evaluated by quotienting out the symmetry factor, namely,
\be
    M_{i,j} = \frac{2!}{\text{symm. factor}}\,,\qquad \bar M_{i,j} = \frac{4!}{\text{symm. factor}}\,,
\ee
where the factorials $2!$ and $4!$ in the numerator are results of the insertion of the quadratic and quartic operators respectively. For instance, the symmetry factor for $D_{0,1}$ is $2!$, due to the pair of the dotted lines, while for $D
_{1,1}$, it is $2!2!$, as there are two pairs. The multiplication factors for all the diagrams under consideration are given as,
\begin{align}
    \begin{aligned}
        M_{0,1} & = M_{0,2} = 1\,,~~~ M_{1,1}  =\frac{1}{2}\,,~~~ M_{2,0} =  \frac{1}{6}\,,\\
        \bar M_{1,0} &=1\,, ~~~ \bar M_{2,0} =  \frac{3}{2}\,,~~~ \bar M_{1,1} = 4\,,~~~ \bar M_{0,2} = 3\,.
    \end{aligned}
\end{align}

The integrals $I_{i,j}$ and $\bar I_{i,j}$ can be evaluated  using the free bulk propagator,
\be
    G(x,y) 
        =
            \int \frac{\d^d p}{(2\pi)^{d}} \frac{e^{\i\, p(x-y)}}{p^2}
        = 
            \frac{C_{\phi\phi}}{|x-y|^{d-2}}\,,
    \qquad
    C_{\phi\phi} 
        = 
            \frac{\Gamma(d/2-1)}{4 \pi ^{d/2}}\,,
\ee
with $C_{\phi\phi}$ being the two-point function normalization of the free bulk field. Before giving the explicit formulas of those integrals, let us give an overview of their divergent structures. For the contributions to $\langle\, \phi^2(x)\, \rangle$, only $I_{0,2}(x)$ and $I_{1,1}(x)$ are divergent as  $1/\eps$. This means that to one-loop order,  $\delta g$ receives corrections from the renormalized surface coupling $h$, as $D_{1,1}$ is a mixed diagram. However, for the contributions to $\langle\, \phi^4(x) \,\rangle$, only $\bar I_{2,0}(x)$ is divergent as $O(1/\eps)$, such that to one-loop order, $\delta h $ is uninfluenced by the renormalized line coupling $g$.
This suggests that the surface coupling has  a higher hierarchy  over the line coupling.

The relevant integrals for the corrections to the line defect coupling are $I_{0,1}(x)$, $I_{0,2}(x)$, $I_{1,1}(x)$, while to the surface coupling, the relevant ones are $\bar I_{1,0}(x)$ and $\bar I_{2,0}(x)$. The results for those integrals are given as following which can be obtained using formulas listed in \eqref{eq:formulaOne}, \eqref{eq:formulaTwo} and \eqref{eq:formulaSan}
of appendix \ref{app:UseFor}
\begin{align}\label{eq:int21}
    I_{0,1}(x)  & = \frac{\pi ^{\frac{1}{2}-d} \,\Gamma \left(\frac{d}{2}-1\right)^2
           \Gamma \left(d-\frac{5}{2}\right)
              }
             {16\, \Gamma (d-2) \left(x_\parallel^2+x_\perp^2\right)^{d-\frac{5}{2}}}
            \,,
    \\ \label{eq:int22}
    I_{0,2}(x) &=
          \frac{(4 \pi) ^{\frac{3}{2} (1-d)}\, \Gamma
        \left(\frac{3}{2}-\frac{d}{2}\right) \Gamma
        \left(d-\frac{5}{2}\right) \Gamma \left(\frac{3
        d}{2}-4\right)
        }{\Gamma \left(\frac{d-1}{2}\right)  \left(x_\parallel^2+x_\perp^2\right)^{\frac{3
        d}{2}-4}}
        \,,
    \\
        I_{1,1} (x) 
   &=\frac{\pi ^{1-2 d}\, \Gamma (3-d) \Gamma \left(\frac{d}{2}-1\right)^4 \Gamma
   \left(d-\frac{5}{2}\right) \Gamma \left(2 d-\frac{11}{2}\right)}{256\, \Gamma (d-2)^2 |x_\perp|^{4d-11}}
   \,,
    \\
    \bar I_{1,0}(x)  &=  \frac{\pi ^{1-2 d}\, \Gamma
   \left(\frac{d-2}{2}\right)^4 \Gamma (2 d-5)}{256\,   \Gamma (2d-4)  |x_\perp|^{4d - 10}}\,,
    \\ \label{eq:int42}
    \bar I_{2,0}(x) 
    &= \frac{\pi ^{2-3 d}\,  \Gamma (3-d) \Gamma
   \left(\frac{d-2}{2}\right)^6 \Gamma (2 d-5)^2 \Gamma (3
   d-8)}
   {4096\, \Gamma (d-2)^2 \Gamma (4 d-10)|x_\perp|^{6 d-16}}\,.
\end{align}


\subsection{Localized RG flows for the single scalar case}
By requiring the finiteness of the one-point functions,  we can obtain the one-loop corrections to the surface and line couplings.  
Collecting the relevant pieces, the requirement of  
the order $O(1/\eps)$ terms to vanish in $\langle\, \phi^4(x)\, \rangle$ 
gives the one-loop correction to the surface coupling
\be
\delta h = \frac{3 h^2}{32\pi} + \text{higher orders} \,.
\ee
It is worth noting that taking the parallel coordinate $x_\parallel = 0$ makes it easier to obtain the above. By using the fact that the bare coupling $h_0$ is independent of the renormalization scale $M$, we can obtain the beta function 
\begin{align}\label{eq:betah}
\beta_h &= M\frac{\d h}{\d M}
= -2\eps h - 2 \delta h +2 h \frac{\der \delta h}{\der h} + \cdots 
= -2\eps h + \frac{3 h^2}{16\pi}  +\cdots\,.
\end{align}
Similarly for $\langle\, \phi^2(x) \,\rangle$, collecting relevant pieces gives the one-loop correction to the line coupling and then the beta function
\begin{align}
\delta g &= \frac{16 g^2+g h}{32 \pi } +  \text{higher orders}\,,\\
\label{eq:betag1}
    \beta_ g &= -\eps g -\delta g -\frac{\beta_h}{\eps} \frac{\der \delta g}{\der h} + g \frac{\der \delta g}{\der g} +\cdots
    =-\eps g + \frac{8g^2+ gh}{16\pi}  +\cdots \,,
\end{align}
where the renormalized surface coupling $h$ enters into the one-loop corrections of the line coupling. 
\begin{figure}[t]
  \centering
  \includegraphics[width=12cm]{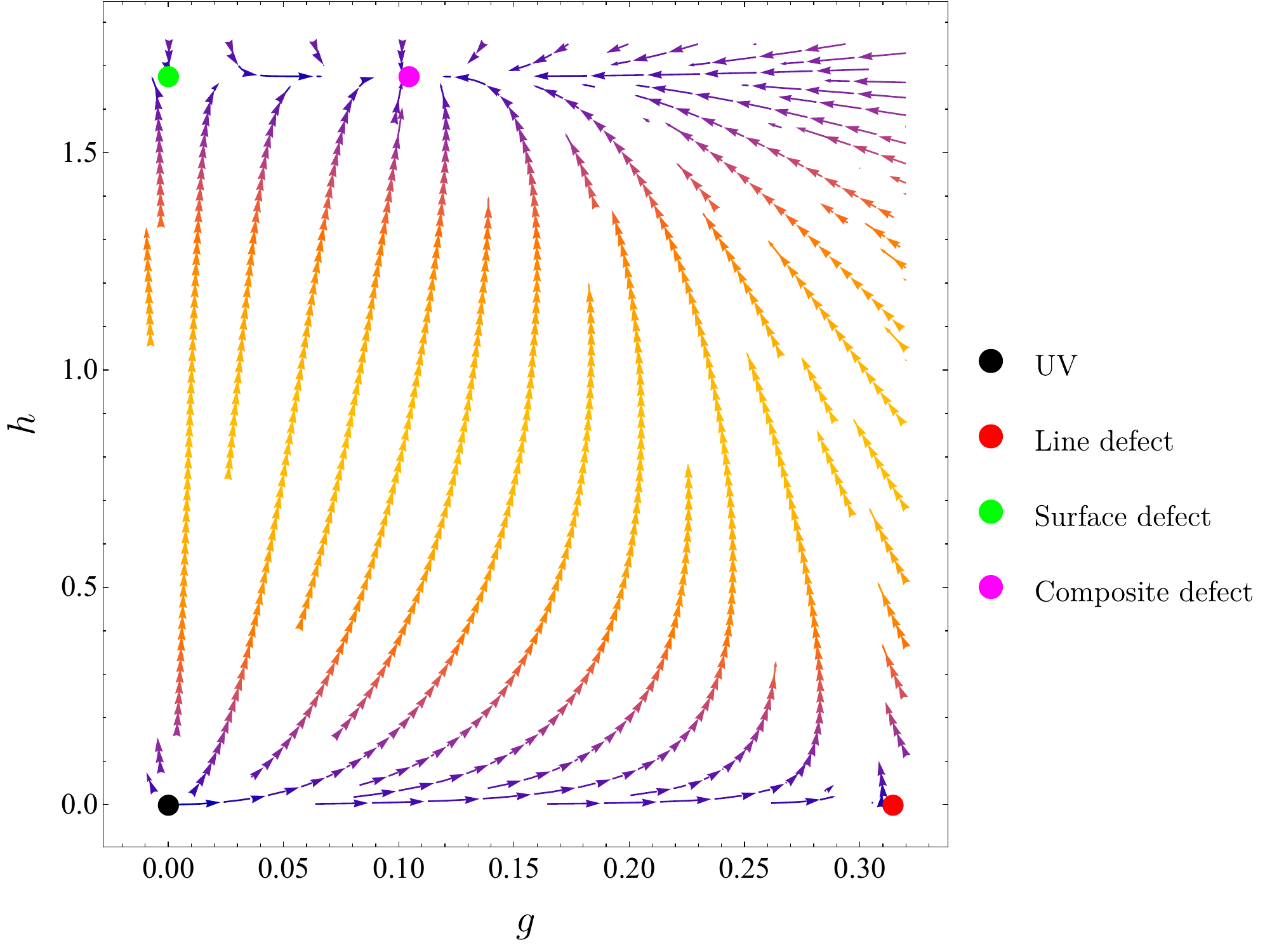}
  \caption{RG flows plotted for $\eps = 0.05$. The black point represents the UV fixed point, the red/green point represents the non-trivial IR fixed point with only the line/surface defect, while the magenta point represents the non-trivial IR fixed point with the composite defect. }
\label{fig:RGflowSingleS}
\end{figure}
The beta functions $\beta_h$ and $\beta_g$ determine the RG flows as plotted in figure \ref{fig:RGflowSingleS}. When both couplings are turned off, the theory is trivial and has the UV fixed point (black); when either the line or the surface coupling is turned on, the UV theory flows to a non-trivial IR fixed point (red or green); most interestingly, when both couplings are turned on, there exists a non-trivial IR fixed point (magenta), which realizes the composite defect at criticality. In the current situation with a single scalar theory, this composite defect fixed point is the most infrared among the three IR fixed points.
Requiring both beta functions \eqref{eq:betah} and \eqref{eq:betag1} to vanish gives the explicit expression for the
critical surface and line defect couplings:
\begin{align}\label{eq:fixedpoint_Ising}
    h_* &= \frac{32\pi}{3}\eps +O(\epsilon^2)\,,
    \\ \label{eq:fixedpoint_Ising2}
    g_* &= 2\pi \eps - \frac{h_*}{8} +O(\epsilon^{2}) = \frac{2\pi}{3}\eps +O(\epsilon^2) \,.
\end{align}
Notice that if we set $\epsilon=0$, then the composite defect CFT does not exist. This observation is consistent with the fact that there is no non-trivial line or surface defects in a single free scalar theory for three-dimentional bulk \cite{Lauria:2020emq}.
The existence of the critical composite defects is highly non-trivial, even though our discussion is mainly on the free bulk theory. As we show in appendix \ref{app:higerD}, such a composite defect fixed point does not exist for all the higher dimensional $d>4$ free scalar theories. This captures the essence of our main result; in the next section we  extend the studies to the $\O(N)$ case.


\section{O$(N)$ vector model with composite defects}\label{sec:ONVecMod}
In the previous section, we obtained the conformal fixed point which realizes the composite defect CFT in a free scalar theory by focusing on the single scalar case. In this section, we extend the analysis  to the O($N$) vector model, and explore various features of the model.


\subsection{Localized RG flows and analysis of fixed points}\label{subsec:FP_analysis}
\begin{figure}[t]
    \centering
    \includegraphics[width=12cm]{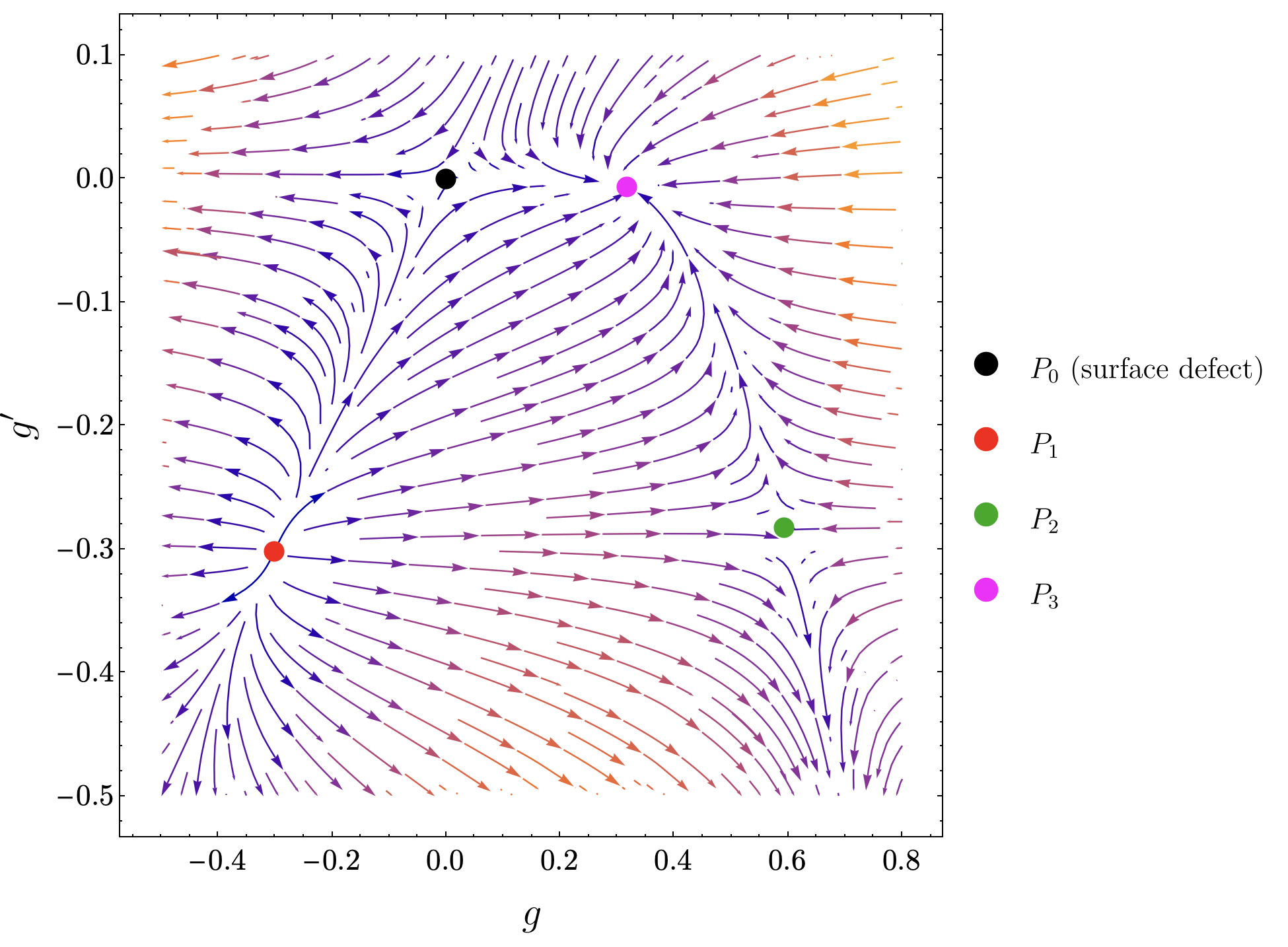}
    \caption{Sub-defect RG flow of the composite defect model in the case of $(N, m, \epsilon)=(300, 2, 0.05)$. Black point corresponds to the trivial theory where a conventional surface defect CFT is realized. The other three points are non-trivial conformal fixed points where composite defect CFTs are emergent. Red point ($P_{1}$) means the composite defect CFT where the O$(300)$ symmetry is preserved, while green and magenta points ($P_{2}$ and $P_{3}$) are ones where such a global symmetry is broken to its sub-group O$(298)\times$O(2) on the sub-defect. There exist some RG flows between O(300) symmetry preserving and breaking phases.}
    \label{fig: RG flow unitary}
\end{figure}

Clearly, we have a lot of symmetry breaking patterns on both the surface and line defects unlike the single scalar case. For simplicity, we pay our attention to the case where the O$(N)$ symmetry is preserved on the surface, but not necessary on the line. Then, the action of our model in $d=3-\epsilon$ can be written as
\begin{align}\label{eq: action}
    I
        \equiv 
            \frac{1}{2}\int_{\BR^{d}} \d^{d}x\,\partial_{\mu}\phi^I \partial^{\mu}\phi^I +\frac{h_{\text{0}}}{4!}\int_{\BR^{2}} \d^{2}\hat{y}\, \left(\phi^I\phi^I \right)^{2}+\frac{g_{0, IJ}}{2}\int_{\BR} \d \tilde{z}\, \phi^{I}\phi^{J} \ .
\end{align}
Here $\phi^I$ is the O$(N)$ vector field, and $I, J$ run from 1 to $N$. Also, $h_{0}$ and $g_{0, IJ}$ are bare couplings for surface and line actions, respectively. 
In an analogous manner as the previous section, we can compute the beta functions for these couplings at one-loop order,
\begin{align}\label{eq:bet_function}
 \left\{
     \begin{aligned}
         \beta_{h}&=-2\epsilon\, h + \frac{N+8}{48\pi}\, h^2+\text{(higher order)}\ , \\
         \beta_{IJ}&=-\epsilon\, g_{IJ}+\frac{1}{2\pi}(g^{2})_{IJ} +\frac{1}{48\pi}\left(\text{Tr}g\, \delta_{IJ}+2g_{IJ}\right) h +\text{(higher order)}\ , 
     \end{aligned}
     \right.
\end{align}
where $h$ and $g_{IJ}$ are renormalized couplings, the beta functions $\beta_{h}$ and $\beta_{IJ}$ are defined by
 \begin{align}
     \beta_{h}\equiv M\frac{\d h}{\d M}\ , \qquad \beta_{IJ}\equiv M\frac{\d g_{IJ}}{\d M}\  .
 \end{align}
Conformal fixed points are obtained by solving the equations $\beta_{h}=\beta_{IJ}=0$. The former condition leads to the critical surface coupling $h_{*}$ as follows
\begin{align}
\label{eq:fixed_point_h}
    h_{*}=\frac{96\,\pi}{N+8}\,\epsilon + O(\epsilon^2)\ . 
\end{align}
Substituting this critical value to the equation $\beta_{IJ}=0$,  it turns out that critical sub-defect couplings $(g_{*})_{IJ}$ have to satisfy the following matrix quadratic equations
\begin{align}\label{eq: quadratic equation}
    (N+8)(g_{*}^{2})_{IJ}-2(N+4)\pi \epsilon (g_{*})_{IJ}+4\pi\epsilon\,  \text{Tr}g_{*}\, \delta_{IJ}=0\ . 
\end{align}
From this equation, we can immediately see that $\text{Tr}g_{*}\not=0$ for any real symmetric matrix $(g_{*})_{IJ}$. 
We can easily prove this statement by contradiction. Suppose $\text{Tr}(g_{*})=0$. By taking trace with respect to O$(N)$ indices in the quadratic equation \eqref{eq: quadratic equation}, we then obtain $\text{Tr} (g_{*}^2)=0$ in general. Since we now assume that $g_{*}$ is a real symmetric matrix, this result implies that only trivial solutions are possible. This completes the proof. Solving this quadratic equations \eqref{eq: quadratic equation} is, however, a quite hard problem, hence we proceed our analysis by paying our attention to the following symmetry breaking patterns on the sub-defect: 
O$(N)\rightarrow \ $O($m$)$\times$O($N-m$). In this case, the coupling constants can be written as
    \begin{align}\label{eq:pattern2}
        \begin{aligned}
            g_{IJ}=\text{diag}\, (\, \underbrace{g\, ,\, g\, , \cdots, \, g}_{m}\, ,\,  \underbrace{g'\, ,\, g'\, , \cdots, \, g'}_{N-m}\, )\, , 
        \end{aligned}
    \end{align}
and the beta functions become
    \begin{align}\label{eq:betag}
        \beta_{g}&=-\epsilon\, g +\frac{1}{2\pi}g^{2}+\frac{1}{48\pi}(m\, g +(N-m)g'+2g)h+\cdots , \\
        \label{eq:betagp}
        \beta_{g'}&=-\epsilon\, g' +\frac{1}{2\pi}{g'}^{2}+\frac{1}{48\pi}(m\, g +(N-m)g'+2g')h+\cdots . 
    \end{align}
Notice that without loss of generality, we can set $m\leq N/2$. Under this restriction, the quadratic equation \eqref{eq: quadratic equation} is reduced to the following simultaneous equations
    \begin{align}
        \begin{aligned}
            (N+8)g^2 +4(N-m)\pi\epsilon\, g'-2 (N-2m+4)\pi\epsilon\, g &=0\ , \\
            (N+8){g'}^2 +2(N-2m-4)\pi\epsilon\, {g'} +4m\pi\epsilon\, g&=0\ .
        \end{aligned}
    \end{align}
In addition to the trivial solution:
    \begin{align}\label{eq:conformal_sub-defect_coupling0}
        P_{0} \ :\  g_{*}= g_{*}'=0\ , 
    \end{align}
there are three non-trivial solutions which are given by
    \begin{align}\label{eq:conformal_sub-defect_coupling1}
    \begin{aligned}
        P_{1} \ :\  g_{*}= g_{*}'=\frac{8-2N}{N+8}\pi\epsilon + O(\epsilon^2)\ , 
    \end{aligned}
    \end{align}
    \begin{align}\label{eq:conformal_sub-defect_coupling2}
    P_{2} \ : \ \left\{
    \begin{aligned}
        g_{*}&=\frac{3N-4m+4+\sqrt{D_{N, m}}}{N+8}\pi\epsilon+O(\epsilon^2)\ ,  \\
         g_{*}'&=\frac{-N+4m+4-\sqrt{D_{N, m}}}{N+8}\pi\epsilon +O(\epsilon^2)\ , 
    \end{aligned}
    \right.
    \end{align}
    \begin{align}\label{eq:conformal_sub-defect_coupling3}
    P_{3}\ : \ \left\{
    \begin{aligned}
        g_{*}&=\frac{3N-4m+4-\sqrt{D_{N, m}}}{N+8}\pi\epsilon+ O(\epsilon^2)\ ,  \\
         g_{*}'&=\frac{-N+4m+4+\sqrt{D_{N, m}}}{N+8}\pi\epsilon+ O(\epsilon^2)\ , 
    \end{aligned}
    \right.
    \end{align}
where $D_{N, m}$ is the discriminant defined by
    \begin{align}
        D_{N, m}\equiv N^2 +16m^2 -16Nm -8N +16\ . 
    \end{align}
Unlike $P_{0}$ and $P_{1}$, two fixed points $P_{2}$ and $P_{3}$ can be possibly complex depending on the sign of $D_{N, m}$. Indeed, we can easily check that for $N\leq 22$, these two fixed points become complex CFTs regardless of the value of $m$. Also, when $m=N/2$ for any even integer $N$, $P_{2}$ and $P_{3}$ again describe complex CFTs. It is not until $N\geq 23$ that unitary CFTs appear for the following values of $m$:
 \begin{align}
 \label{eq:specific_m_unitary}
     m=1\,, \ 2\, , \cdots ,\, \left\lfloor{\frac{N}{2}-\frac{\sqrt{(3N-4)(N+4)}}{4}}\right\rfloor\ . 
 \end{align}
Below, we will see the behaviors of RG flows around unitary and complex CFTs, respectively.
 \paragraph{Renormalization group flow around unitary CFTs.}
 \begin{figure}[t]
    \begin{center}
        \begin{tabular}{ccc}
        \begin{minipage}{0.3\linewidth}
            \begin{center}
                \begin{tikzpicture}[scale=0.50, samples=300]
                   \coordinate (C) at (0 , 0) {};
                   \coordinate (D) at (8 , 0) {};
                 \foreach \x in {0,2}
                {
                    \draw[thick, blue!50!black, arrows = {-Stealth[scale=1.1]}] ({4+4*cos(180*\x/2)}, {4*sin(180*\x/2)})--({4+4*cos(180*\x/2)/2}, {4*sin(180*\x/2)/2});
                    \draw[thick, blue!50!black] ({4+4*cos(180*\x/2)/2}, {4*sin(180*\x/2)/2}) -- ($(C)!0.5!(D)$);
                }
                 \foreach \x in {1,3}
                {
                    \draw[thick, blue!50!black, arrows = {-Stealth[scale=1.1]}] ($(C)!0.5!(D)$) -- ({4+4*cos(180*\x/2)/2}, {4*sin(180*\x/2)/2});
                    \draw[thick, blue!50!black] ({4+4*cos(180*\x/2)/2}, {4*sin(180*\x/2)/2}) -- ({4+4*cos(180*\x/2)}, {4*sin(180*\x/2)});
                }
                \draw[thick, domain=8:4.55, blue!50!black, arrows = {-Stealth[scale=1.1]}] plot({\x}, {1/(\x-4)});
                \draw[thick, domain=4.55:4.25, blue!50!black] plot({\x}, {1/(\x-4)});
                \draw[thick, domain=8:4.55, blue!50!black, arrows = {-Stealth[scale=1.1]}] plot({\x}, {-1/(\x-4)});
                \draw[thick, domain=4.55:4.25, blue!50!black] plot({\x}, {-1/(\x-4)});
                \draw[thick, domain=8:4.55, blue!50!black, arrows = {-Stealth[scale=1.1]}] plot({-\x+8}, {1/(\x-4)});
                \draw[thick, domain=4.55:4.25, blue!50!black] plot({-\x+8}, {1/(\x-4)});
                \draw[thick, domain=8:4.55, blue!50!black, arrows = {-Stealth[scale=1.1]}] plot({-\x+8}, {-1/(\x-4)});
                \draw[thick, domain=4.55:4.25, blue!50!black] plot({-\x+8}, {-1/(\x-4)});
        \node at ($(C)!0.5!(D)+(0, -5)$) {$\lambda_{+}\lambda_{-}<0$};
        \node at ($(C)!0.5!(D)+(0, -6.5)$) {type 1};
         \filldraw[fill=white] ($(C)!0.5!(D)$) circle (0.15);
         \fill[red!50] ($(C)!0.5!(D)$) circle (6pt);
                \end{tikzpicture}    
                \end{center}
            \end{minipage}
            &
            \begin{minipage}{0.3\linewidth}
            \begin{center}
               \begin{tikzpicture}[scale=0.50]
                   \coordinate (C) at (0 , 0) {};
                   \coordinate (D) at (8 , 0) {};
        \foreach \x in {0,1,2, 3}
                {
                    \draw[thick, blue!50!black, arrows = {-Stealth[scale=1.1]}] ({4+4*cos(180*\x/2)}, {4*sin(180*\x/2)})--({4+4*cos(180*\x/2)/2}, {4*sin(180*\x/2)/2});
                    \draw[thick, blue!50!black] ({4+4*cos(180*\x/2)/2}, {4*sin(180*\x/2)/2}) -- ($(C)!0.5!(D)$);
                }
        \node at ($(C)!0.5!(D)+(0, -5)$) {$\lambda_{+}>0$, $\lambda_{-}>0$};
        \node at ($(C)!0.5!(D)+(0, -6.5)$) {type 2};
         \filldraw[fill=white] ($(C)!0.5!(D)$) circle (0.15);
         \fill[red!50] ($(C)!0.5!(D)$) circle (6pt);
                \end{tikzpicture}    
                \end{center}
            \end{minipage}
            &
            \begin{minipage}{0.3\linewidth}
            \begin{center}
                \begin{tikzpicture}[scale=0.50]
                   \coordinate (C) at (0 , 0) {};
                   \coordinate (D) at (8 , 0) {};
        \foreach \x in {0,1,2, 3}
                {
                    \draw[thick, blue!50!black, arrows = {-Stealth[scale=1.1]}] ($(C)!0.5!(D)$) -- ({4+4*cos(180*\x/2)/2}, {4*sin(180*\x/2)/2});
                    \draw[thick, blue!50!black] ({4+4*cos(180*\x/2)/2}, {4*sin(180*\x/2)/2}) -- ({4+4*cos(180*\x/2)}, {4*sin(180*\x/2)});
                }
        \node at ($(C)!0.5!(D)+(0, -5)$) {$\lambda_{+}<0$, $\lambda_{-}<0$};
        \node at ($(C)!0.5!(D)+(0, -6.5)$) {type 3};
         \filldraw[fill=white] ($(C)!0.5!(D)$) circle (0.15);
         \fill[red!50] ($(C)!0.5!(D)$) circle (6pt);
                \end{tikzpicture}    
                \end{center}
            \end{minipage}
        \end{tabular}
        \caption{Classification of local RG flows around the fixed points (depicted as red points) in the $(g_{2}, g_{2}')$ plane. The blue lines mean the RG flows, and the arrows lying on them point in the IR direction. The local RG behaviors depend on the signs of eigenvalues $\lambda_{\pm}$ of the Jacobi matrix $J$.}
        \label{fig:classification_local_RG_flow}
    \end{center}
\end{figure}
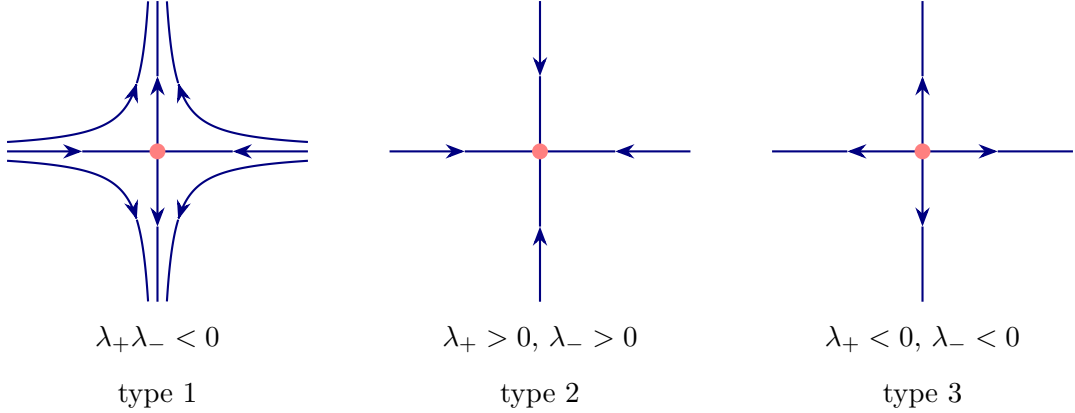
We first discuss the sub-defect RG flow in our composite defect system by focusing on the case where both fixed points $P_{2}$ and $P_{3}$ are unitary (namely, the discriminant $D_{N, m}$ is positive).   
To analyze the stability around the conformal fixed points, we can perturb the sub-defect couplings around the fixed points:
 \begin{align}
     g=g_{*}+\delta g\ , \qquad g'=g'_{*}+\delta g'\ ,
 \end{align}
with the surface defect coupling fixed to be $h_{*}$. By plugging these expressions into \eqref{eq:bet_function}, we can then obtain the following linear differential equations with respect to $\delta g_{2}$ and $\delta g_{2}'$:
\begin{align}
    \begin{aligned}
        \frac{\d \delta \bm{g}}{\d\log M}=J\,\delta \bm{g}\ , \qquad \delta \bm{g}\equiv \begin{pmatrix}
            \delta g \\
            \delta g'
        \end{pmatrix} \ ,  
    \end{aligned}
\end{align}
where $J$ is the Jacobi matrix with the following form
\begin{align}
    \begin{aligned}
        J=\epsilon
        \begin{pmatrix}
            -1+\frac{g_{*}}{\pi \epsilon}+\frac{2m+4}{N+8} & \frac{2N-2m}{N+8} \\
            \frac{2m}{N+8} & -1+\frac{g'_{*}}{\pi \epsilon}+\frac{2N-2m+4}{N+8}
        \end{pmatrix}
        +O(\epsilon^{2})\ . 
    \end{aligned}
\end{align}
As described in figure \ref{fig:classification_local_RG_flow}, the local behaviors of RG flows around the fixed points can be completely characterized by the eigenvalues $\lambda_{\pm}$ of the above Jacobi matrix $J$, given as
\begin{align}
    \begin{aligned}
        \lambda_{\pm}(g_{*}\, ,\, g_{*}')=\frac{(N+8)(g_{*}+g_{*}')-8\pi\epsilon\pm\sqrt{D}}{2(N+8)\pi}\ , 
    \end{aligned}
\end{align}
where $D$ is defined by
\begin{align}
    D\equiv(N+8)^{2}(g_{*}-g_{*}')^{2}-4(N+8)(N-2m)(g_{*}-g_{*}')\pi\epsilon +4N^{2}\pi^{2}\epsilon^{2}\ . 
\end{align}
By substituting the explicit values of fixed points \eqref{eq:conformal_sub-defect_coupling0}\,--\,\eqref{eq:conformal_sub-defect_coupling3} to the eigenvalues $\lambda_{\pm}$, we obtain 
\begin{align}
    \begin{aligned}
        P_{0}\ : \ \lambda_{+}=\frac{N-4}{N+8}\epsilon\ , \qquad \lambda_{-}=-\frac{N+4}{N+8}\epsilon\ , 
    \end{aligned}
\end{align}
\begin{align}
    \begin{aligned}
        P_{1}\ : \ \lambda_{+}=-\frac{N-4}{N+8}\epsilon\ , \qquad \lambda_{-}=-\frac{3N-4}{N+8}\epsilon\ , 
    \end{aligned}
\end{align}
\begin{align}
    P_{2}\ : \ \lambda_{\pm}=\frac{N\pm \sqrt{2}\sqrt{N^{2}+8m^{2}-8Nm-4N+8+(N-2m)\sqrt{D_{N, m}}}}{N+8}\epsilon\ , 
\end{align}
\begin{align}
    P_{3}\ : \ \lambda_{\pm}=\frac{N\pm \sqrt{2}\sqrt{N^{2}+8m^{2}-8Nm-4N+8-(N-2m)\sqrt{D_{N, m}}}}{N+8}\epsilon\ .
\end{align}
Notice that the eigenvalues for $P_{2}$ and $P_{3}$ are all real for unitary theories, namely, $N\geq23$ and $m$ is in the range of \eqref{eq:specific_m_unitary}, although it might be somewhat non-trivial. From these expressions, it turns out that the local RG flows around $P_{0}$ and $P_{2}$ belong to the type 1 while local flows around $P_{3}$ ($P_{1}$) do to the type 2 (type 3), respectively in figure \ref{fig:classification_local_RG_flow}. 
 In figure \ref{fig: RG flow unitary}, we also plot the the RG flow in the $(g, g')$-plane with the fixed surface defect coupling $h=h_{*}$ and setting $(N, m, \epsilon)=(300, 2, 0.05)$. Interestingly, this figure implies that there exist non-trivial RG flows between O($N$) symmetry preserved and breaking phases. In section \ref{sec:subdefecttest}, we see that this obtained RG flow is completely consistent with the sub-defect $\CC$-theorem.

\begin{figure}[t]
     \centering
      \includegraphics[width=8cm]{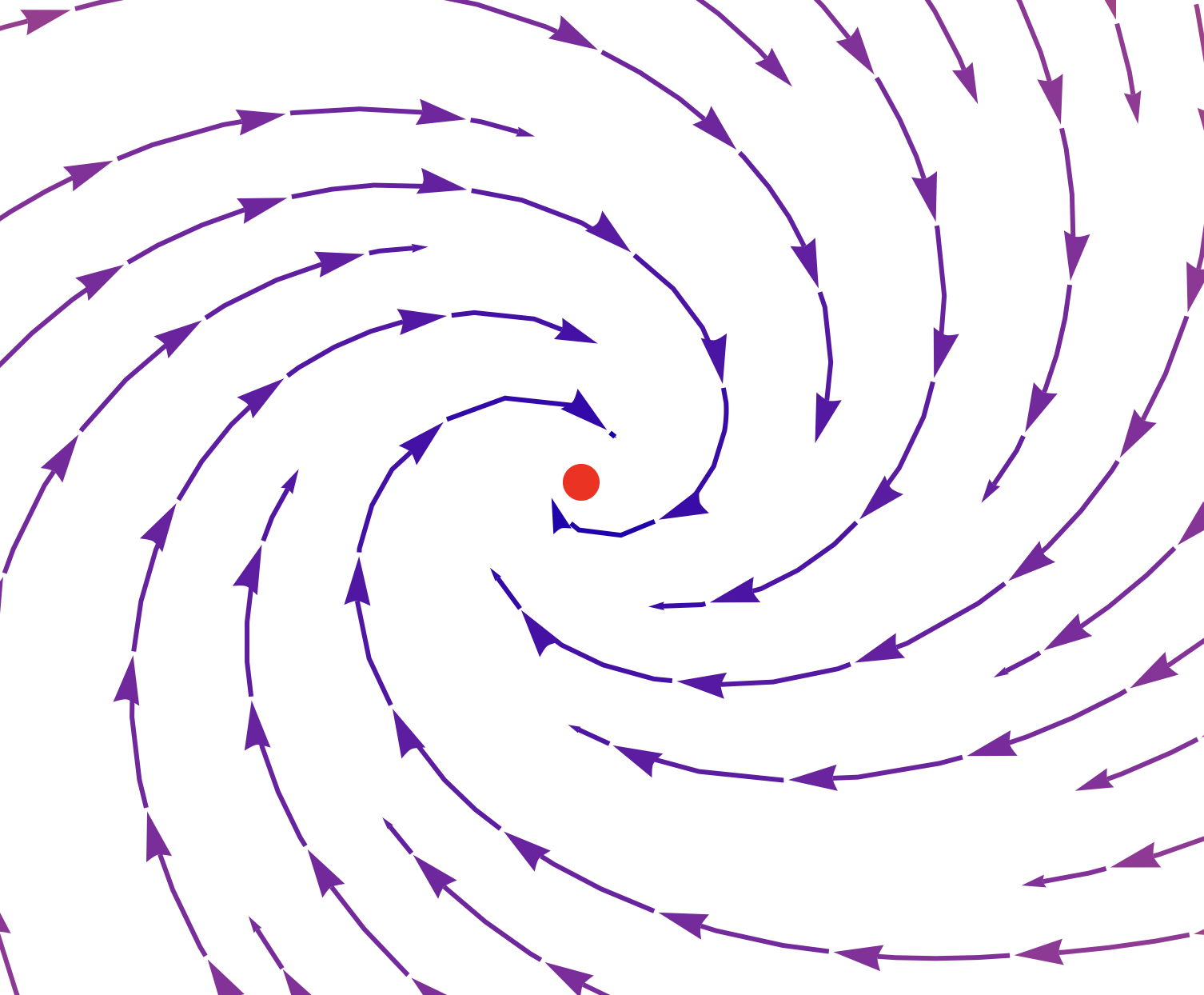}
     \caption{The localized RG flow around the complex CFT point in the $(\slg, \tilde{\slg})$-plane. The red point represents the complex fixed point \eqref{eq: complex FP}, and the arrow on the RG flow points to the IR direction. This fixed point is IR stable at least at the leading order.}
     \label{fig: RG flow complex cft}
 \end{figure}

\paragraph{Renormalization group flow around complex CFTs.} 
As described previously, when $m=N/2$ for arbitrary even integer $N$, the fixed points $P_{2}$ and $P_{3}$ become complex numbers. In particular, one of them is given by
\begin{align}\label{eq: complex FP}
    \frac{g_{*}}{\pi\epsilon}=\frac{N+4}{N+8}+\i\frac{\sqrt{(3N-4)(N+4)}}{N+8}\ , \qquad \frac{g_{*}'}{\pi\epsilon}=\frac{N+4}{N+8}-\i\frac{\sqrt{(3N-4)(N+4)}}{N+8}\ .
\end{align}
We next explore the local RG flow around this complex CFT point. For this purpose, we promote the originally real couplings $g$ and $g'$ to the complex ones as follows:
\begin{align}
    g\equiv \slg+\i\, \tilde{\slg}\qquad , \qquad g'\equiv \slg'+\i\, \tilde{\slg}'\ ,  
\end{align}
where $\slg, \tilde{\slg}, \slg'$ and $\tilde{\slg}'$ are all real couplings. Correspondingly, we can define the beta functions for each coupling as follows:
\begin{align}
        \beta_{\slg}\equiv \frac{\d \slg}{\d \log M}\ , \quad \beta_{\tilde{\slg}}\equiv \frac{\d \tilde{\slg}}{\d \log M}\ , \quad  \beta_{\slg'}\equiv \frac{\d \slg'}{\d \log M}\ , \quad \beta_{\tilde{\slg}'}\equiv \frac{\d \tilde{\slg}'}{\d \log M}\ . 
\end{align}
We now have the four-dimensional coupling space, and it is hard to visualize RG flows on such a space. Therefore, in the following, we completely fix the values of $\slg'$ and $\tilde{\slg}'$ to be the critical points \eqref{eq: complex FP}, namely we set 
\begin{align}
    \slg'=\slg_{*}'=\frac{N+4}{N+8}\,\pi\epsilon\ , \qquad \tilde{\slg}'=\tilde{\slg}_{*}'=-\frac{\sqrt{(3N-4)(N+4)}}{N+8}\,\pi\epsilon\ , 
\end{align}
and discuss on the two-dimensional $(\slg, \tilde{\slg})$-plane. To explore RG behaviors at a little away region from the complex CFT point \eqref{eq: complex FP}, we perturb the two couplings $\slg$ and $\tilde{\slg}$ in the following way:
\begin{align}
    \slg=\slg_{*}+\delta \slg\ , \qquad   \tilde{\slg}=\tilde{\slg}_{*}+\delta \tilde{\slg}\ , 
\end{align}
where $\slg_{*}$ and $\tilde{\slg}_{*}$ are the fixed point \eqref{eq: complex FP} defined by
\begin{align}
     \slg_{*}=\frac{N+4}{N+8}\,\pi\epsilon\ , \qquad \tilde{\slg}_{*}=\frac{\sqrt{(3N-4)(N+4)}}{N+8}\,\pi\epsilon\ . 
\end{align}
By using the beta functions \eqref{eq:betag} and \eqref{eq:betagp}, we can obtain the following differential equations up to leading term:
\begin{align}
    \begin{aligned}
        \frac{\d \delta \bm{\slg}}{\d\log M}=\mathsf{J}\,\delta \bm{\slg}\ , \qquad \delta \bm{\slg}\equiv 
        \begin{pmatrix}
            \delta \slg \\
            \delta \tilde{\slg}
        \end{pmatrix} \ ,  
    \end{aligned}
\end{align}
where $\mathsf{J}$ is the Jacobi matrix which is defined by
\begin{align}
    \begin{aligned}
        \mathsf{J}=\frac{\epsilon}{N+8}
        \begin{pmatrix}
        N & -\sqrt{(3N-4)(N+4)} \\
        \sqrt{(3N-4)(N+4)} & N
        \end{pmatrix}
        +O(\epsilon^{2})\ . 
    \end{aligned}
\end{align}
As easily checked. unlike the localized RG flows around the real fixed points, the eigenvalues of the matrix $\mathsf{J}$ are complex-valued for $N\geq 2$. We also notice that the real parts of these eigenvalues are positive. This implies that the complex CFT point \eqref{eq: complex FP} is IR stable in the $(\slg, \tilde{\slg})$-plane, which is illustrated in figure \ref{fig: RG flow complex cft}. 
Interestingly, this RG behavior can be seen as the linear approximation of the equation (41) in the literature \cite{Jepsen:2020czw}, and we naturally expect that the higher order contributions to the beta function produce the chaotic RG behavior. We leave more detailed investigations on RG flows around complex CFT points to future studies.


\subsection{Anomalous dimensions of defect and sub-defect local primaries}
In the previous section, we investigated the RG structures of composite defect systems, and succeeded in finding conformal fixed points which realize the composite defect CFTs. Composite defects at the critical point can be characterized by conformal data e.g., anomalous dimensions. In this section, hence, we derive the anomalous dimensions of sub-defect local primaries. 
\subsubsection{Anomalous dimensions of O$(N)$ fundamental fields}
\begin{figure}[t]
  \centering
  \begin{tikzpicture}[thick]
      \begin{scope}[xshift=0cm]
          \draw[-] (-2, 0) -- (2, 0);
          \node[cross=4pt] at (-1.75, 0) {};
          \node[cross=4pt] at (1.75, 0) {};
          \draw[dotted, bend left=60] (-1.75, 0) to (1.75, 0);
      \end{scope}

      \begin{scope}[xshift=6cm]
          \draw[-] (-2, 0) -- (2, 0);
          \node[cross=4pt] at (-1.75, 0) {};
          \node[cross=4pt] at (1.75, 0) {};
          \node[dot=8pt] at (0, 0) {};
          \draw[dotted, bend left=60] (-1.75, 0) to (0, 0);
          \draw[dotted, bend left=60] (0, 0) to (1.75, 0);
      \end{scope}
  \end{tikzpicture}
  \caption{Feynman diagrams contributing to the two-point function of sub-defect fundamental fields.
  The dashed line represents the propagator and the cross symbols represent the sub-defect fundamental field.
  Also, the circle represents the interaction localized on the line defect.}
\label{fig:anomaloud_dimension_fund}
\end{figure}
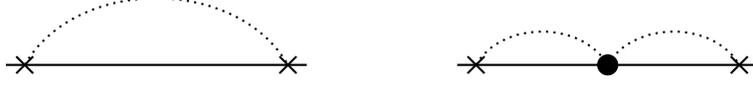
As a warm-up, we start with the anomalous dimension of the sub-defect fundamental O$(N)$ vector field $\phi^I(\tilde{z})$. To compute it, we must calculate the sub-defect two-point function $\langle\, {\phi}^I(\tilde{z})\, {\phi}^J(0) \, \rangle$ at the one-loop level. This evaluation can be performed by using the perturbative method as before (see figure \ref{fig:anomaloud_dimension_fund}):
\begin{align}\label{eq:integral_pert1}
    \begin{aligned}
        \langle\, {\phi}^{I}(\tilde{z})\, {\phi}^{J}(0) \, \rangle 
        &=\langle\, {\phi}^{I}(\tilde{z})\, {\phi}^{J}(0) \, \rangle_{0}-g_{0, KL}\int_{\BR}\d \tilde{w} \langle\, {\phi}^{I}(\tilde{z})\, {\phi}^{K}(\tilde{w})\, \rangle_{0} \langle\, {\phi}^{J}(0)\,\, {\phi}^{L}(\tilde{w})\, \rangle_{0} \\
        &=\frac{C_{\phi\phi}\, \delta^{IJ}}{|\tilde{z}|^{1-\epsilon}}-\frac{g_{0, IJ}}{4\pi^2}\left(\frac{1}{\epsilon}+\aleph+O(\epsilon)\right)\frac{1}{|\tilde{z}|^{1-2\epsilon}}\ , 
    \end{aligned}
\end{align}
where $\aleph$ is numerical constant which is defined by
\begin{align}
    \aleph\equiv \gamma_{\text{E}}+\log 4\pi\ . 
\end{align}
Below, we focus on the symmetry breaking case $\text{O}(N)\to \text{O}(m)\times \text{O}(N-m)$ (see \eqref{eq:pattern2}). We introduce the wavefunction renormalization factors $Z_{m}$ and $Z_{N-m}$:
\begin{align}
    \begin{aligned}
    \phi^{\alpha}(\tilde{z})&=Z_{m}\, [\,\phi^{\alpha}\,](\tilde{z})\ \ , \qquad \  \alpha=1\sim m\ , \\
    \phi^{i}(\tilde{z})&=Z_{N-m}\, [\,\phi^{i}\,](\tilde{z})\ ,  \qquad i=m+1\sim N \ , 
    \end{aligned}
\end{align}
where $[\,\phi^{\alpha}\,](\tilde{z})$ and $[\,\phi^{i}\,](\tilde{z})$ are renormalized sub-defect local operators. We choose the factors $Z_{m}$ and $Z_{N-m}$ such that the pole appearing in \eqref{eq:integral_pert1} is cancelled:
\begin{align}\label{eq:wavefunc_ren_fund}
    Z_{m}\cong 1-\frac{g}{2\pi}\frac{1}{\epsilon}\ , \qquad Z_{N-m}\cong 1-\frac{g'}{2\pi}\frac{1}{\epsilon}\ .
\end{align}
From this, we can compute the anomalous dimensions of the fundamental sub-defect operators at the critical point:
\begin{align}\label{eq:anomaldimphi}
    \gamma_{m}\equiv \left.\frac{\d \log Z_{m}}{\d \log M}\right|_{\text{fixed point}}\cong\frac{g_{*}}{2\pi}\ , \qquad \gamma_{N-m}\equiv \left.\frac{\d \log Z_{N-m}}{\d \log M}\right|_{\text{fixed point}}\cong\frac{g'_{*}}{2\pi}\ . 
\end{align}
We can then arrive at the following explicit forms of the sub-defect two-point functions:
\begin{align}
    \langle\, [\,{\phi}^{\alpha}\,](\tilde{z})\, [\,{\phi}^{\beta}\, ](0) \, \rangle=\delta^{\alpha\beta}\frac{\CN_{m}}{|\tilde{z}|^{2\widetilde{\Delta}_m}}\ , \qquad \CN_{m}\equiv C_{\phi\phi}\left(1-\frac{g_{*}}{2\pi}\aleph+O(\epsilon^2 )\right) \ , 
\end{align}
\begin{align}
    \langle\, [\,{\phi}^{i}\,](\tilde{z})\, [\,{\phi}^{j}\, ](0) \, \rangle=\delta^{ij}\frac{\CN_{N-m}}{|\tilde{z}|^{2\widetilde{\Delta}_{N-m}}}\ , \qquad \CN_{N-m}\equiv C_{\phi\phi}\left(1-\frac{g'_{*}}{2\pi}\aleph+O(\epsilon^2 )\right) \ , 
\end{align}
where $\widetilde{\Delta}_m$ and $\widetilde{\Delta}_{N-m}$ are conformal dimensions of renormalized fundamental operators $[\,{\phi}^{\alpha}\,](\tilde{z})$ and $[\,{\phi}^{j}\,](\tilde{z})$, respectively:
\begin{align}
    \widetilde{\Delta}_m\cong 1-\frac{\epsilon}{2}+\frac{g_{*}}{2\pi}+O(\epsilon^2)\ , \qquad \widetilde{\Delta}_{N-m}\cong 1-\frac{\epsilon}{2}+\frac{g'_{*}}{2\pi}+O(\epsilon^2)\ . 
\end{align}
\subsubsection{Anomalous dimensions of sub-defect composite operators}
We next move on to the anomalous dimensions of sub-defect composite operators.
In particular, we concentrate on the following scalar primaries with the canonical scaling dimension one:
\begin{align}
\label{eq:def_phi_psi}
    \Phi\equiv \frac{1}{\sqrt{m}}\sum_{\alpha=1}^{m}\,(\phi^{\alpha})^2\ , \qquad \Psi\equiv \frac{1}{\sqrt{N-m}}\sum_{i=m+1}^{N}(\phi^{i})^2\ ,
\end{align}
which are the interacting terms localized on the line defect \eqref{eq: action}. Also, the normalization factors are chosen for later convenience. We remark that these operators have the same canonical dimensions, hence we can expect that $\Phi$ and $\Psi$ should mix to each other due to the quantum corrections. 
The renormalized fields $[\Phi]$ and $[\Psi]$ are related to the bare fields $\Phi$ and $\Psi$ as
\begin{align}\label{eq:mixing_renormalization}
\begin{aligned}
\begin{pmatrix}
        \Phi \\
        \Psi
\end{pmatrix}
= Z_{S}
\begin{pmatrix}
    [\Phi] \\
    [\Psi]
\end{pmatrix}
\end{aligned}\ , 
\end{align}
where $Z_{S}$ is the wavefunction renormalization matrix whose size is two. By using this matrix, we can compute the anomalous dimension matrix $\gamma$ which is defined by
\begin{align}\label{eq:anomalous_dimension_matrix}
    \begin{aligned}
        \gamma\equiv \left. Z_{S}^{-1}\frac{\d}{\d \log M} Z_{S}\right|_{\text{fixed point}}
    \end{aligned}
\end{align} 
From the general grounds, we can write this matrix in the following manner:
\begin{align}
    Z_{S}=\bm{1}_{2\times 2}+\delta  Z_{S}\ , 
\end{align}
where $\bm{1}_{2\times 2}$ is the unit matrix. By requiring that the following three-point functions:
\begin{align}\label{eq:three_pt_func}
    \begin{aligned}
    \langle\, [\,{\phi}^{I}\,](\tilde{z}_{1})\, [\,{\phi}^{J}\,](\tilde{z}_{2})\, [\Phi](0) \, \rangle\ , \qquad \langle\, [\,{\phi}^{I}\,](\tilde{z}_{1})\, [\,{\phi}^{J}\,](\tilde{z}_{2})\, [\Psi](0) \, \rangle\ , 
    \end{aligned}
\end{align}
should have no poles, we can fix the renormalization matrix as follows:
\begin{align}\label{eq:renormalization_matrix}
    \begin{aligned}
     \delta  Z_{S}=-\frac{1}{96\pi \epsilon}
        \begin{pmatrix}
            (m+2)h+96g & \sqrt{m(N-m)}h \\
            \sqrt{m(N-m)}h & (N-m+2)h+96g'
        \end{pmatrix} \ .
    \end{aligned}
\end{align}
(See appendix \ref{app:computaton_anomalous_dim} for more details.)
By plugging this into \eqref{eq:anomalous_dimension_matrix}, we can obtain the anomalous dimension matrix at the leading order:
\begin{align}
\label{eq:gamma_matrix}
    \begin{aligned}
        \gamma &=\frac{1}{48\pi}
        \begin{pmatrix}
            (m+2)h_{*}+48g_{*} & \sqrt{m(N-m)}h_{*} \\
            \sqrt{m(N-m)}h_{*} & (N-m+2)h_{*}+48g_{*}'
        \end{pmatrix}+O(\epsilon^{2})\ , 
    \end{aligned}
\end{align}
where $h_{*}$, $g_{*}$ and $g_{*}'$ are fixed points which are given in section \ref{subsec:FP_analysis}. Notice that the above anomalous dimension matrix is symmetric. This is the reason why we introduced the factors $\sqrt{m}$ and $\sqrt{N-m}$ in \eqref{eq:def_phi_psi}.\footnote{Equivalently, one can introduce these factors into the renormalization matrix factors instead of including these factors in the definitions of $\Phi$ and $\Psi$.} The eigenvalues $\gamma_{\pm}$ of this matrix gives rise to the anomalous dimensions of the two operators $S_{\pm}$ resulting from the mixing of operators $\Phi$ and $\Psi$, which are given by
\begin{align}\label{eq:anomalgamma}
    \gamma_{\pm}=\frac{48(g_{*}+g'_{*})+(N+4)h_{*}\pm\sqrt{192\Delta_{g} (12\Delta_{g}+m h_{*})-96N\Delta_{g}h_{*}+N^{2}h_{*}^{2}}}{96\pi}\ , 
\end{align}
where $\Delta_{g}\equiv g_{*}-g'_{*}$. One notes that for the $\O(N)$ preserving case, $\Delta_g = 0$, the anomalous dimensions for the two operators are independent on the partition $m$. Substituting the fixed point values for  $P_1$, the anomalous dimensions gives
\be
\gamma_\pm\Big|_{P_1} = \frac{12-N \pm N}{8+N}\eps\,.
\ee


\section{Towards a $\mathcal{C}$-theorem in composite defect CFTs}\label{sec:sub-defect_C_theorem}

For a theory with a defect $\CD^{(p)}$ of $p$ dimensions, one can trigger an RG flow localized on $\CD^{(p)}$.
In this circumstance, the defect $\CC$-theorem is conjectured to hold, which states the existence of a function called the $\CC$-function that decreases under any localized RG flow \cite{Kobayashi:2018lil}.
Motivated by the defect $\CC$-theorem conjecture, we extend it to the case with a composite defect and propose the following:

\noindent
\begin{shaded}\noindent
\textbf{Conjecture.}
In a unitary CFT$_d$ with a composite defect $\CD^{(p_1,\,\cdots,\,p_n)} = \cup_{i=1}^n\,\CD^{(p_i)}$ consisting of $n$ sub-defects of $p_i$ dimensions satisfying $0 < p_1 < p_2 < \cdots < p_n < d$ and $\CD^{(p_1)} \subset \CD^{(p_2)} \subset \cdots \subset \CD^{(p_n)}$, let $Z^{(p_1,\,\cdots,\,p_n)} \equiv \langle\,\CD^{(p_1,\,\cdots,\,p_n)}\,\rangle$ be the partition function on a $d$-sphere.
Then, the function $\CC$ defined by
\begin{align}\label{sub-defect-C-function}
    \CC 
        \equiv 
        \sin\left(\frac{\pi p_1}{2}\right)\, \log\,\left| \frac{Z^{(p_1,\, p_2 \, , \,\cdots,\,p_n)}}{Z^{(p_2,\,\cdots,\,p_n)}}\right| 
\end{align}
does not increase under any localized RG flow on the sub-defect $\CD^{(p_1)}$ of the lowest dimension,
\begin{align}\label{weak_C}
    \CC_\mathrm{UV} \ge \CC_\mathrm{IR} \ .
\end{align}
\end{shaded}
Note that our conjecture is a weak $\CC$-theorem which compares the values of the sub-defect $\CC$-function \eqref{sub-defect-C-function} at the UV and IR fixed points connected by a localized RG flow.
While the $\CC$-function \eqref{sub-defect-C-function} is defined for the theories away from the fixed points, it may not be monotonically decreasing function under the flow, hence be a weak $\CC$-function.\footnote{There are three types of $\CC$-functions: weak, strong, and strongest ones \cite{Barnes:2004jj,Nishioka:2018khk}.
The function $\CC$ satisfying \eqref{weak_C} is called the weak $\CC$-function while it is called the strong $\CC$-function when $\CC$ is a monotonically decreasing function under an RG flow.
The function $\CC$ is called the strongest $\CC$-function if an RG flow is its gradient flow.
}

\subsection{Conformal perturbation theory on sub-defect}
We examine the validity of our conjecture in the conformal perturbation theory with slightly relevant perturbation.
The following argument is a slight modification of the previous works in \cite{Cardy:1988cwa,Klebanov:2011gs,Nozaki:2012qd,Gaiotto:2014gha,Kobayashi:2018lil}.

Let $I^{(p_1,\,\cdots,\,p_n)}_\text{DCFT}$ be the action of a CFT with a composite defect $\CD^{(p_1,\,\cdots,\,p_n)}$, and consider an RG flow localized on the sub-defect $\CD^{(p_1)}$ of lowest dimensions:
\begin{align}\label{perturbed_action}
    I^{(p_1,\,\cdots,\,p_n)}
        = 
            I^{(p_1,\,\cdots,\,p_n)}_\text{DCFT} 
                + 
                \tilde{\lambda}_0\,\int \d^{p_1} \tilde{x}\, \sqrt{\tilde{ g}}\,\widetilde{\CO}(\tilde{x}) \ ,
\end{align}
where $\widetilde\CO(\tilde x)$ is a slightly relevant primary operator of dimension $\widetilde \Delta = p_i - \varepsilon$ with small $\varepsilon$ on $\CD^{(p_1)}$, and $\tilde\lambda_0$ is the bare coupling at the UV cutoff scale $\mu_\text{UV}$.\footnote{We use $\varepsilon$ for the slightly relevant perturbations to distinguish from $\epsilon$ for the $\epsilon$-expansion.}
The two-point function at the unperturbed DCFT is normalized on $\BR^d$ as 
\begin{align}\label{two-point-function}
    \langle\, \widetilde\CO(\tilde x_1)\,\widetilde\CO(\tilde x_2)\,\rangle_0
        =
        \frac{1}{|\tilde x_1 - \tilde x_2|^{2\widetilde \Delta}} \ ,
\end{align}
while the three-point function is fixed by the residual conformal symmetry on $\CD^{(p_1)}$ as\footnote{$\langle\, \widetilde\CO(\tilde x)\,\rangle_0$ stands for the vev of the local operator $\widetilde\CO(\tilde x)$ at the unperturbed DCFT.}
\begin{align}\label{three-point-function}
    \langle\, \widetilde\CO(\tilde x_1)\,\widetilde\CO(\tilde x_2)\,\widetilde\CO(\tilde x_3)\,\rangle_0
        =
        \frac{\widetilde C}{|\tilde x_1 - \tilde x_2|^{\widetilde \Delta}\,|\tilde x_2 - \tilde x_3|^{\widetilde \Delta}\,|\tilde x_3 - \tilde x_1|^{\widetilde \Delta}} \ ,
\end{align}
where $\widetilde C$ is a constant.
It follows from the correlation functions that the OPE of a pair of the local operator $\widetilde\CO$ is
\begin{align}
    \widetilde\CO(\tilde x_1)\,\widetilde\CO(\tilde x_2)
        \underset{x_1\to x_2}{=}
             \frac{1}{|\tilde x_1 - \tilde x_2|^{2\widetilde \Delta}}
             +
              \widetilde C\,\frac{\widetilde\CO(\tilde x_1)}{|\tilde x_1 - \tilde x_2|^{\widetilde \Delta}}
            + \cdots \ .
\end{align}

Now consider a correlation function $\langle\,\cdots\,\rangle$ of the perturbed DCFT on $\BR^d$ and expand it as a function of $\tilde\lambda_0$ as
\begin{align}
    \begin{aligned}
        \langle\,\cdots\,\rangle
            &=
                \langle\,\cdots\,\rangle_0
                -
                \tilde\lambda_0\,\int \d^{p_1}\tilde x_1\,\langle\,\widetilde\CO(\tilde x_1)\,\cdots\,\rangle_0 \\         
            &\qquad\quad +
                \frac{\tilde\lambda_0^2}{2}\,\int \d^{p_1}\tilde x_1\int \d^{p_1}\tilde x_2\,\langle\,\widetilde\CO(\tilde x_1)\,\widetilde\CO(\tilde x_2)\,\cdots\,\rangle_0
                + \cdots \\
            &=
                \langle\,\cdots\,\rangle_0
                -
                \tilde\lambda_0\,\int \d^{p_1}\tilde x_1\,\langle\,\widetilde\CO(\tilde x_1)\,\cdots\,\rangle_0 \\   
            &\qquad +
                \frac{\tilde\lambda_0^2}{2}\,\int \d^{p_1}\tilde x_1\int \d^{p_1}\tilde x_2\,\frac{\widetilde C}{|\tilde x_1 - \tilde x_2|^{\widetilde\Delta}}\langle\,\widetilde\CO(\tilde x_1)\,\cdots\,\rangle_0
                + \cdots \\
            &=
                \langle\,\cdots\,\rangle_0
                -
                \tilde\lambda(\mu)\,\int \d^{p_1}\tilde x_1\,\langle\,\widetilde\CO(\tilde x_1)\,\cdots\,\rangle_0
                + \cdots
                \\   
    \end{aligned}
\end{align}
where $\tilde\lambda(\mu)$ is the effective coupling at the energy scale $\mu$:
\begin{align}
    \begin{aligned}
    \tilde\lambda(\mu)
        &=
            \tilde\lambda_0 
                - \frac{\tilde\lambda_0^2}{2}\,\int_{\mu_\text{UV}^{-1} < |\tilde x_1 - \tilde x_2| < \mu^{-1}} \d^{p_1}\tilde x_2\,\frac{\widetilde C}{|\tilde x_1 - \tilde x_2|^{\widetilde\Delta}} + \cdots \\
        &=
            \tilde\lambda_0 
                - 
                \tilde\lambda_0^2\,\frac{\pi^\frac{p_1}{2}\,\widetilde C}{\varepsilon\,\Gamma\left(\frac{p_1}{2}\right)}\left[ \mu^{-\varepsilon} - \mu_\text{UV}^{-\varepsilon} \right] + \cdots \ .
    \end{aligned}
\end{align}
By introducing the dimensionless couplings $\tilde g(\mu) = \tilde\lambda(\mu)\,\mu^{-\varepsilon}$ and $\tilde g_0 = \tilde\lambda_0\,\mu_\text{UV}^{-\varepsilon}$, we obtain
\begin{align}
    \tilde g(\mu)
        =
        \tilde g_0\,\left( \frac{\mu_\text{UV}}{\mu}\right)^\varepsilon 
                    - 
                    \tilde g_0^2\,\frac{\pi^\frac{p_1}{2}\,\widetilde C}{\varepsilon\,\Gamma\left(\frac{p_1}{2}\right)}\left[ \left( \frac{\mu_\text{UV}}{\mu}\right)^{2\varepsilon} - \left( \frac{\mu_\text{UV}}{\mu}\right)^\varepsilon \right] + \cdots \ .
\end{align}
The beta function for the dimensionless coupling becomes
\begin{align}
    \beta(\tilde g)
        =
            \mu\,\frac{\d \tilde g}{\d \mu}
        =
            -\varepsilon\,\tilde g 
            +
            \frac{\pi^\frac{p_1}{2}\,\widetilde C}{\Gamma\left(\frac{p_1}{2}\right)}\,\tilde g^2 + \cdots \ ,
\end{align}
which, if $\widetilde C >0$, has the IR fixed point at
\begin{align}\label{CPT_IR_fixed_point}
    \tilde g_\ast 
        = 
            \frac{\Gamma\left(\frac{p_1}{2}\right)}{\pi^\frac{p_1}{2}\,\widetilde C}\,\varepsilon + O(\varepsilon^2) \ .
\end{align}

 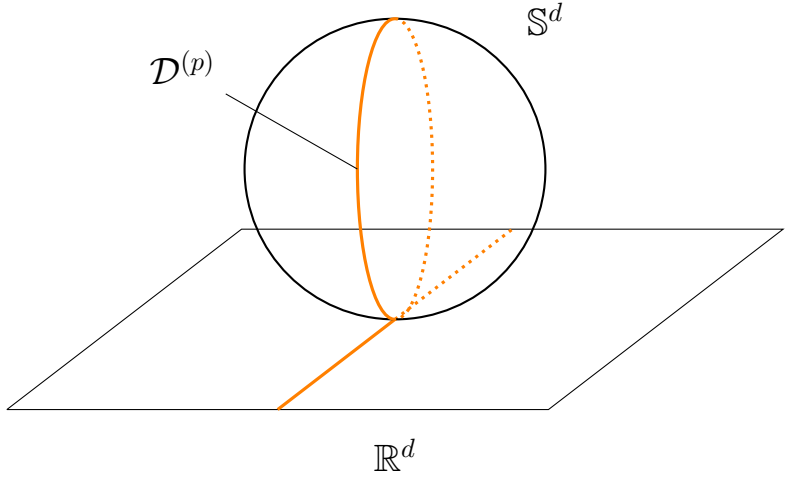
\begin{figure}[t]
  \centering
  \begin{tikzpicture}
      \begin{scope}[scale=1.2] 
          \draw[-] (-4.3, -1) -- (1.7, -1) -- (4.3, 1) -- (-1.7, 1) -- (-4.3, -1);
          \draw[very thick, orange] (-1.3, -1) -- (0,0);      
          \draw[very thick, orange, dotted] (0, 0) -- (1.3, 1);
          \node at (0, -1.5) {\Large $\BR^d$};
      \end{scope}

      \begin{scope}[yshift=2cm]
        \draw[thick] (0,0) circle (2cm);
        \draw[-, very thick, orange] ($(0, 0) + (90:0.5cm and 2cm)$(P) arc (90:270:0.5cm and 2cm);
        \draw[-, very thick, orange, dotted] ($(0, 0) + (90:0.5cm and 2cm)$(P) arc (90:-90:0.5cm and 2cm);
        \node at (2, 2) {\Large $\BS^d$};
        \draw[-] ($(0, 0) + (180:0.5cm and 2cm)$(P) -- (-2.25, 1) node[above left] {\Large $\CD^{(p)}$};
      \end{scope}
  \end{tikzpicture}
  \caption{The stereographic projection of $\BS^d$ to $\BR^d$. A $p$-dimensional spherical defect $\CD^{(p)}$ is mapped to a planar defect on $\BR^d$.}
\label{fig:stereographic}
\end{figure}

By mapping the perturbed DCFT \eqref{perturbed_action} on $\BR^d$ to a $d$-sphere $\BS^d$ of radius $R$ (see figure \eqref{fig:stereographic}) and expanding the sphere partition function $Z^{(p_1,\,\cdots,\,p_n)}\left(\tilde\lambda_0\right)$around $\tilde\lambda_0 = 0$, we obtain
\begin{align}\label{log_Z}
    \begin{aligned}
        \log\,Z^{(p_1,\,\cdots,\,p_n)}\left(\tilde\lambda_0\right)
            &=
                \log\,Z^{(p_1,\,\cdots,\,p_n)}(0) 
                +
                \frac{\tilde\lambda_0^2}{2}\,\CI_2 - \frac{\tilde\lambda_0^3}{6}\,\CI_3 + O\left(\tilde\lambda_0^4\right) \ ,
    \end{aligned}
\end{align}
where $\CI_2$ and $\CI_3$ are the two- and three-point connected functions in the unperturbed theory integrated over the sub-defect $\CD^{(p_1)}$,
\begin{align}
        \CI_2
            &=
                \int \d^{p_1}\tilde x_1\sqrt{\tilde{ g}}\int \d^{p_1}\tilde x_2\sqrt{\tilde{ g}}\,\langle\,\widetilde\CO(\tilde x_1)\,\widetilde\CO(\tilde x_2) \,\rangle_0 \ , \label{I_2}\\
        \CI_3
            &=
                \int \d^{p_1}\tilde x_1\sqrt{\tilde{ g}}\int \d^{p_1}\tilde x_2\sqrt{\tilde{ g}}\int \d^{p_1}\tilde x_3\sqrt{\tilde{ g}}\,\langle\,\widetilde\CO(\tilde x_1)\,\widetilde\CO(\tilde x_2) \,\widetilde\CO(\tilde x_3)\,\rangle_0 \label{I_3}\ .      
\end{align}
To evaluate $\CI_2$ and $\CI_3$, it is convenient to use the stereographic coordinates of the $d$-sphere:
\begin{align}
    \d s^2_{\BS^d}
        =
        (2R)^2\,\frac{\sum_{\mu=1}^d(\d x^\mu)^2}{\left( 1 + |x|^2\right)^2} \ .
\end{align}
The stereographic projection can be seen as a conformal mapping from $\BR^d$ to $\BS^d$, where a planar defect $\CD^{(p_1)}$ located at $x^{i} = 0~(i = p_1+1, \cdots, d)$ is mapped to a spherical defect of radius $R$.
The two- and three-point functions after the conformal map take the same forms as \eqref{two-point-function} and \eqref{three-point-function}, respectively, with $|\tilde x_1 - \tilde x_2|$ replaced with the chordal distance $s(\tilde x_1, \tilde x_2) = 2R\,|\tilde x_1 - \tilde x_2|/\sqrt{\left( 1 + |\tilde x_1|^2\right)\left( 1 + |\tilde x_2|^2\right)}$ on $\BS^{p_1}$.
Substituting them into \eqref{I_2} and \eqref{I_3} and performing the integration on $\BS^{p_1}$ of radius $R$, we obtain
\begin{align}
    \begin{aligned}
        \CI_2
            &=
            2^{2\varepsilon} \pi^{p_1} R^{2\varepsilon}\,\frac{\Gamma\left(\varepsilon - \frac{p_1}{2}\right)\,\Gamma\left(\frac{p_1}{2}\right)}{\Gamma\left(p_1\right)\,\Gamma(\varepsilon)} 
            =
            -\frac{2^{2\varepsilon +1} \pi^{p_1+1} R^{2\varepsilon}}{\sin\left(\frac{\pi p_1}{2}\right)\,\Gamma(p_1+1)}\,\frac{\Gamma\left(\varepsilon - \frac{p_1}{2}\right)}{\Gamma\left(-\frac{p_1}{2}\right)\,\Gamma(\varepsilon)} 
            \ , \\
        \CI_3
            &=
            8\,\pi^{\frac{3(p_1 +1)}{2}}R^{3\varepsilon}\,\frac{\Gamma\left( \frac{3\varepsilon - p_1}{2}\right)}{\Gamma(p_1)\,\Gamma\left(\frac{\varepsilon+1}{2}\right)^3} \,\tilde C\ .
    \end{aligned}
\end{align}
On the other hand, the bare dimensionless couplings are related by
\begin{align}
    \tilde\lambda_0\,(2R)^\varepsilon 
        =
        \tilde g + \frac{\pi^\frac{p_1}{2}\,\widetilde C}{\varepsilon\,\Gamma\left(\frac{p_1}{2}\right)}\,\tilde g^2 + O(\tilde g) \ ,
\end{align}
where $\tilde g = \tilde g(\mu)$ and we choose  the energy scale as $\mu = 1/(2R)$  while letting $\mu_\text{UV} \to \infty$ \cite{Klebanov:2011gs}.
With this relation, it follows from \eqref{log_Z} that the deviation of the sphere partition function from that at the UV fixed point ($\tilde g =0$) is
\begin{align}
    \begin{aligned}
        \delta\log\,Z^{(p_1,\,\cdots,\,p_n)}\left(\tilde g\right)
            &\equiv
            \log\,Z^{(p_1,\,\cdots,\,p_n)}\left(\tilde g\right)
            -
            \log\,Z^{(p_1,\,\cdots,\,p_n)}\left(0\right) \\
            &=
            \frac{2\,\pi^{p_1 + 1}}{\sin\left(\frac{\pi p_1}{2}\right)\,\Gamma(p_1+1)}
            \left[
                - \frac{\varepsilon}{2}\,\tilde g^2 + \frac{\pi^\frac{p_1}{2}\,\tilde C}{3\,\Gamma\left(\frac{p_1}{2}\right)}\, \tilde g^3 + O(\tilde g^4)
            \right]\ .
    \end{aligned}
\end{align}
Substituting the IR value of the dimensionless coupling \eqref{CPT_IR_fixed_point} and recalling the definition of the $\CC$-function \eqref{sub-defect-C-function}, we find
\begin{align}\label{eq:difference}
    \CC(\tilde g_\ast) - \CC(\tilde g =0)
        =
        -\frac{\pi\,\Gamma\left(\frac{p_1}{2}\right)^2}{3\,\Gamma(p_1+1)}\,\frac{\varepsilon^3}{\tilde C^2} + O(\varepsilon^4) \ ,
\end{align}
which is negative as is consistent with our sub-defect $\CC$-theorem conjecture.


\subsection{Perturbative test in $\O(N)$ vector model with composite defects}\label{sec:subdefecttest}
In the last subsection, we gave a general argument for the validity of the conjectured sub-defect $\CC$-theorem in terms of the conformal perturbation theory with a one-parameter deformation.
In the current subsection, we examine that by considering the  $\O(N)$ vector model studied in section \ref{sec:ONVecMod}, where a line defect is embedded inside a surface defect, \ie  $p_2=2$ and $p_1=1$.
In particular, we study the case where two line deformations which preserves the $\O(m)\times \O(N-m)$ internal symmetry  are turned on, giving rise to three non-trivial IR fixed points. Using the usual perturbative approach, we can compute the $\CC$-functions (or the modified free energy) at all the IR fixed points. Adapting the definition of the $\CC$-function, in \eqref{sub-defect-C-function}, we have 
\be\label{eq:free energy}
    \CC = \ln \frac{Z(h_0,g_0,g_0')}{ Z(h_0,0,0)} =  C_2 + C_3 + C_3' + (\text{higher order}) \ ,
\ee
where we keep the terms up to third order in the couplings,  $C_2$ and $C_3$ denote the quadratic and cubic terms in the line couplings, while $C_3'$ denotes the mixed term between the surface and the line coupling. The contributions purely from the surface defect is subtracted from the partition function in the denominator with only the surface defect turned on.  

 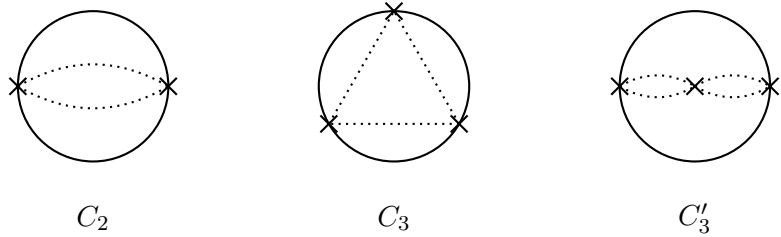
\begin{figure}[h!]
  \centering

  \begin{tikzpicture}[thick]
      \begin{scope}
          \draw (0,0) circle (1cm);
          \node[cross=4pt] at (-1, 0) {};
          \node[cross=4pt] at (1, 0) {};
          \draw[dotted, bend right] (-1, 0) to (1, 0);
          \draw[dotted, bend left] (-1, 0) to (1, 0);
          \node at (0, -1.75) {$C_2$};
      \end{scope}

      \begin{scope}[xshift=4cm]
          \draw (0,0) circle (1cm);
          \node[cross=4pt] at (-30:1) {};
          \node[cross=4pt] at (90:1) {};
          \node[cross=4pt] at (210:1) {};
          \draw[dotted] (-30:1) to (90:1);
          \draw[dotted] (-30:1) to (210:1);
          \draw[dotted] (210:1) to (90:1);
          \node at (0, -1.75) {$C_3$};
      \end{scope}

      \begin{scope}[xshift=8cm]
          \draw (0,0) circle (1cm);
          \node[cross=4pt] at (-1, 0) {};
          \node[cross=4pt] at (1, 0) {};
          \node[cross=4pt] at (0, 0) {};
          \draw[dotted, bend right] (-1, 0) to (0, 0);
          \draw[dotted, bend left] (-1, 0) to (0, 0);
          \draw[dotted, bend left] (0, 0) to (1, 0);
          \draw[dotted, bend right] (0, 0) to (1, 0);
          \node at (0, -1.75) {$C_3'$};
      \end{scope}
  \end{tikzpicture}
  \caption{The three diagrams for the contributions to the sub-defect $\CC$ function up to the third order of the couplings, $C_2$ and $C_3$ denote the quadratic and cubic terms   in the line couplings, while $C_3'$ denotes the mixed term between the surface and the line coupling. }
\label{fig:free_energy}
\end{figure}
Normally, we evaluate the partition function on $\BS^d$ to get rid of the IR divergences. However, the ratio between the partition functions gives the same result if we map $\BS^d$ to $\mathbb{R}^d$, as the IR divergences and the conformal anomaly induced by the surface defect cancel. In order to understand the localized RG flow on the line defect, a further map is required for the evaluation. 
The line defect is mapped to a circle $\BS^1$ of radius $R$ while the surface defect is kept as the plane $\mathbb{R}^2$, then separately each term gives
\begin{align}
    C_2 
        &= 
        \frac{C_{\phi\phi}^2}{4}\,
        (m\, g_0^2 + (N-m)\, g_0'^2)\, I_2\,, \\
    C_3 
        &= 
        -\frac{C_{\phi\phi}^3}{6}\,(m\, g_0^3 + (N-m)\, g_0'^3)\, I_3\,, \\
    C_3' 
        &= 
        - \frac{C_{\phi\phi}^4}{4!}\, h_0 \left[m(m+2)\, g_0^2 + (N-m)(N-m+2)\, g_0'^2 + 2m(N-m) \,g_0\, g_0'\right]\,I_3' \,,
\end{align}
where $I_2, I_3, I_3'$ are given by
\begin{align}
    I_2 
        &= 
        \int \d^2 z_1\, \d^2 z_2\, \frac{ \delta(|z_1|-R)\, \delta(|z_2|-R)}{|z_1 - z_2|^{2(d-2)}} 
        = 
        4^{\eps}\,\pi^\frac{3}{2}\, R^{2\eps} \,\frac{\Gamma \left(\eps-\frac{1}{2}\right)}{\Gamma(\eps)}\,,\\
    I_3 
        &= 
        \int \d^2 z_1\, \d^2 z_2\, \d^2 z_3 \, 
        \frac{ \delta(|z_1|-R)\, \delta(|z_2|-R)\,\delta(|z_3|-R)}{|z_1 - z_2|^{(d-2)}\, |z_1 - z_3|^{(d-2)}\, |z_2 - z_3|^{(d-2)} }  
        =
        8 \pi^3\, R^{3\eps} \,\frac{\Gamma \left(\frac{3\eps -1}{2} \right)}{\Gamma
           \left(\frac{1+\eps}{2}\right)^3}\,,\\
     I_3' 
        & = 
        \int \d^2 z_1\, \d^2 z_2\, \d^2 y\,
        \frac{ \delta(|z_1|-R)\, \delta(|z_2|-R)}{|z_1 -y|^{2(d-2)}\, |z_2 -y|^{2(d-2)} } 
        =
        4^{1-\eps}\,\pi ^3 R^{4\eps} \,\frac{\Gamma \left(2\eps-\frac{1}{2}\right) \Gamma
           \left(\frac{1}{2}-\eps\right)}{\Gamma \left(\frac{1}{2}+\eps\right)^2 \Gamma (1-\eps)}\,,
\end{align}
Gathering all the terms and keeping 
up to third orders in the coupling, we can write the perturbative result for the  sub-defect $\CC$-function using the beta functions:
\begin{align}
    \CC  
        =
        - \frac{m\,g^3 + (N-m)\,g'^3}{192\, \pi } + \frac{m\, g\, \beta_g + (N-m)\, g'\, \beta_{g'}}{32}\,,
\end{align}
where the beta functions $\beta_g$ and $\beta_{g'}$ are defined as in \eqref{eq:betag} and \eqref{eq:betagp} respectively. One can check to this order that the $\CC$-function is stationary at the conformal fixed points in the following sense that the variations with respect to the renormalized couplings are proportional to the corresponding beta functions,
\be
    \frac{\partial \CC}{\partial g} 
        =
            - \frac{m}{16}\, \beta_g\,,
    \qquad
    \frac{\partial \CC}{\partial g'} 
        = 
            - \frac{N-m}{16}\, \beta_{g'}\,.
\ee 
Now we can compare the different fixed points using the modified sub-defect free energy.
For the $\O(N)$ preserving phase, the difference between the trivial $(P_0)$ and non-trivial $(P_1)$ fixed points is
\be
    \CC(P_1) - \CC(P_0) = \frac{\pi^2\, N\,(N-4)^3 }{24\, (N+8)^3}\epsilon ^3 + O(\eps^4)\,,
\ee
of which the leading order is positive for $N>4$. This can be compared with the general analysis in \eqref{eq:difference} using the conformal perturbation theory. For that we need to take $\varepsilon = (4-N)\,\epsilon/(N+8)$ and the three point function (normalized by the two-point function normalization) $\tilde C  = 8N C_{\phi\phi}^3/\left( 2N C_{\phi\phi}^2\right)^{3/2} = \sqrt{8/N}$.
For the DCFT model to be unitary, as put before, $N\ge 23$, this suggests that the non-trivial fixed point is always more UV than the trivial one, which is consistent with the local RG flow analysis. For the $\O(N)$ broken phase, the difference between the two fixed points gives
\be
    \CC(P_2) - \CC(P_3) =\frac{\pi ^2\,  (N-2 m)\,(D_{N,m})^\frac{3}{2}}{24\, (N+8)^3}\,\epsilon ^3 + O(\eps^4)\,,
\ee
which indicates that the UV/IR properties of the two $\O(N)$ fixed points exchanges depending on 
whether $m>N/2$. One can also compare the fixed points between the $\O(N)$ symmetric and $\O(N)$ broken phases, it turns out that the non-trivial $\O(N)$ symmetric fixed point is always the most UV among the four fixed points.


\section{Summary and discussion}\label{sec:conclusion_discussion}

In this work, we have made a first attempt to understand a new class of DCFTs with composite defects, which are to embed a lower-dimensional conformal defect inside a higher-dimensional one.
We find that a concrete example of such DCFTs can be realized by the $\O(N)$ vector model in $d=3-\eps$ dimensions.
Quite interestingly, there exists a composite defect with a line defect embedded into a surface one, even when the bulk theory is free.
This is a distinctive feature from its counterpart at $d=4-\eps$ dimensions where the conformal line defect cannot be realized inside the surface defect.
The reason behind is that the free theory in $d=4-\eps$ is not able to host a non-trivial conformal line defect \cite{Cuomo:2021kfm}. The same reasoning applies for the higher-dimensional cases with $d=2n-\eps $ $(\mathbb{Z} \ni n>2)$ as we have shown in appendix \ref{app:higerD}, though there the lower-dimensional defect is an $(n-1)$-dimensional defect. 
All these suggest that the existence of a composite defect CFT in $d=3-\eps$ is highly non-trivial, thus this dimension has been our main focus of this work.

We initiated our study by considering a single scalar model, whose action is given by \eqref{eq:actiononeboson}.
By turning on  a localized surface $\phi^4$ interaction and a localized line $\phi^2$ interaction, we analyzed the RG flows diagrammatically to the second order of the couplings.
A new Wilson-Fisher fixed point appears in the IR (depicted in figure \ref{fig:RGflowSingleS}), where the line coupling is shifted by the critical value of the surface coupling, as in \eqref{eq:fixedpoint_Ising} and \eqref{eq:fixedpoint_Ising2}.
This is due to the contribution of the mixed diagram $D_{\color{red} 1,1}$ in figure \ref{fig:d3o2diags} to the one-loop correction of the line coupling $g$.
Noting that the critical value of the surface coupling is uninfluenced by the line coupling, which is expected from the point of view of the conformal perturbation theory, for instance as shown in two-dimensional CFTs with boundaries \cite{Gaberdiel:2008fn}.
 
Allowing for $\O(N)$ symmetry breaking terms on the line defect, we further considered the $\O(N)$ vector model with the action \eqref{eq: action}, focusing on a special type of $\O(N)$ symmetry breaking pattern, \ie $\O(N) \to \O(m) \times \O(N-m)$.
Due to the breaking of the internal symmetries, we have two additional non-trivial IR fixed points $P_2$ \eqref{eq:conformal_sub-defect_coupling2} and $P_3$ \eqref{eq:conformal_sub-defect_coupling3}, apart from the $\O(N)$ symmetry preserving ones $P_0$ \eqref{eq:conformal_sub-defect_coupling0} and $P_1$ \eqref{eq:conformal_sub-defect_coupling1}.
Noting that all these IR fixed points locate on the critical plane of the surface coupling with $h=h_*=96\pi \eps/(N+8)$.
When $N\ge 23$, these fixed points describe unitary CFTs.
The $\O(N)$-preserving fixed point $P_1$ is always the most UV among the four, while $P_2$ or $P_3$ is always the most IR depending on whether $m>N/2$ or $m<N/2$ respectively.
On the other hand, when $N< 23$, the $\O(N)$-broken fixed points $P_2$ and $P_3$ describe complex composite defect CFTs since sub-defect couplings become complex-valued. It is well known that in homogeneous systems, the RG flows around complex CFT points typically show different behaviors from the ones around real CFTs \cite{Jepsen:2020czw, Bosschaert:2021ooy}. We observed that by taking $m=N/2$ with $N\in 2\mathbb{Z}^+$, the complex (sub-defect) line  coupling exhibit spiral flow structures at the linear order.

We also considered the operator contents on the line defect of low dimensions.
The fundamental fields on the line defect have anomalous dimensions proportional to the corresponding renormalized couplings, as in \eqref{eq:anomaldimphi} to one-loop order.
They do not receive contributions from the surface coupling at this order.
Interestingly, for the composite operators $\Phi$ and $\Psi$ constructed as sum of quadratic terms in the fundamental fields $\phi^I$,  they get mixed up by the surface coupling at one-loop level to give two operators $S_\pm$ with anomalous dimensions $\gamma_\pm$ as in \eqref{eq:anomalgamma}.
Similar discussions can be found in \cite{Gimenez-Grau:2022czc}.

To capture the property of the submost defect along the localized RG flow, we introduced a $\CC$-function as in \eqref{sub-defect-C-function} for a general composite defect system.
We further conjecture that the $\CC$-function is non-increasing when compared at the UV and IR fixed points connected by a localized RG flow, \ie  $\CC_\text{UV}\ge \CC_\text{IR}$.
We provided evidence for our conjecture by using the conformal perturbation theory.
We also tested it in our $\O(N)$ vector model in $d=3-\eps$ by calculating the $\CC$-function perturbatively and confirmed the validity of our conjecture.

There are a few potentially interesting directions to be explored for the future work. Notably, it is likely that a general proof of the weak $\CC$-theorem on the submost defect exists in dimensions $p_1= 1,2,4$ for a composite DCFT.  
For a local, reflection-positive Euclidean (or unitary Lorentzian) CFT in general dimensions with only one defect $\CD^{(p)}$, the strong $g$-theorem has been proven for a line defect $\CD^{(1)}$ in $d\ge 3$ \cite{Cuomo:2021rkm} by deriving a gradient formula for the defect entropy which decreases monotonically along the defect RG flow\footnote{The $d=2$ case coincides with the proof of the $g$-theorem for the BCFT by Friedan and Konechny \cite{Friedan:2003yc}.};  while the weak version was proved for  a flat surface defect $\CD^{(2)}$ \cite{Jensen:2015swa} and a flat four-volume defect $\CD^{(4)}$ in $d\ge 5$ \cite{Wang:2021mdq}, more in the spirit of \cite{Komargodski:2011vj,Komargodski:2011xv}. 
There the anomaly matching condition was used, to relate the change  (UV to IR) of the $b$-charge to the two-point function of the defect stress tensor, and the change of the $a$-charge to the two-to-two $S$-matrix of the dilatonic scattering, reflection positivity or unitarity ensures the monotonicity of the charges along the defect RG flow.   
One should be able to extend their strategies  \cite{Cuomo:2021rkm,Jensen:2015swa,Wang:2021mdq} to the composite DCFTs.

Another interesting direction, yet unexplored, is the search for other configurations of composite defects in the $\O(N)$ vector model in $d=4-\eps$. 
It has been shown that a volume defect (with real fixed points) exists in the free bulk theory setup for $N>2$ \cite{Harribey:2023xyv} and stable ones put extra constraints on $N$ \cite{Harribey:2024gjn}. 
It would be tempting to examine whether this volume defect can host a line defect inside.
Another possible configuration of composite defects is  a surface defect inside the volume defect.
We leave these studies to future investigations.


\acknowledgments
We are grateful to Pawel Caputa, Sumit Das, Diptarka Das, Nadav Drukker, Damian Galante, Chris Herzog, Miguel Paulos, Massimo Porrati, Ritam Sinha and Maxime Tr\'epanier for valuable discussion. D.\,G. thanks the hospitality of the Theoretical Physics group at King's College London and the University of Warsaw, where this work was presented. The work of T.\,N. was supported in part by the JSPS Grant-in-Aid for Scientific Research (C) No.19K03863, Grant-in-Aid for Scientific Research (B) No.\,24K00629, and Grant-in-Aid for Scientific Research (A) No.\,21H04469.
The works of T.\,N. and D.\,G. were supported in part by the JSPS Grant-in-Aid for Transformative Research Areas (A) ``Extreme Universe'' No.\,21H05182 and No.\,21H05190.
The work of S.\,S. was supported by Grant-in-Aid for JSPS Fellows No.\,23KJ1533.


\appendix

\section{Some useful formulas}\label{app:UseFor}
In this section, we collect a few formulas that are useful for the one-loop calculation,
\begin{align}\label{eq:formulaOne}
    &\int \d^dp\, \frac{p^{2\alpha}}{\left( p^2+M^2 \right)^\beta} = \pi^\frac{d}{2}\,M^{d+2\alpha -2\beta} \frac{\Gamma\left(\alpha + \frac{d}{2}\right) \Gamma\left(\beta - \alpha - \frac{d}{2}\right)}{\Gamma\left(\frac{d}{2}\right)\,\Gamma(\beta)}\,,
    \\
\label{eq:formulaTwo}
    &\int \d^d p\, \frac{1}{\left( k^2 + p^2 \right)^\alpha p^{2\beta}}  = 
    \frac{\pi ^{d/2} \Gamma \left(\frac{d}{2}-\beta \right) \Gamma
       \left(-\frac{d}{2}+\alpha +\beta \right)
       }{\Gamma
       (\alpha ) \Gamma \left(\frac{d}{2}\right)}\left(\frac{1}{k^2}\right)^{\alpha +\beta -\frac{d}{2}}\,,
       \\
\label{eq:formulaSan}
    &\iint \frac{\mathrm{d}^d k \mathrm{~d}^d l }{\left(k^2+m^2\right)^{\alpha}\left[(k+l)^2\right]^{\beta}\left(l^2+m^2\right)^{\gamma}} \no\\
    &\quad =  \frac{\pi^d \Gamma\left(\alpha+\beta-d / 2\right) \Gamma\left(\beta+\gamma-d / 2\right) \Gamma\left(d / 2-\beta\right) \Gamma\left(\alpha+\beta+\gamma-d\right)}{\Gamma\left(\alpha\right) \Gamma\left(\gamma\right) \Gamma\left(\alpha+2 \beta+\gamma-d\right) \Gamma(d / 2)\left(m^2\right)^{\alpha+\beta+\gamma-d}}\,.
\end{align}


\section{Free model in even dimensions $d=2n-\eps$: no composite DCFT}\label{app:higerD}
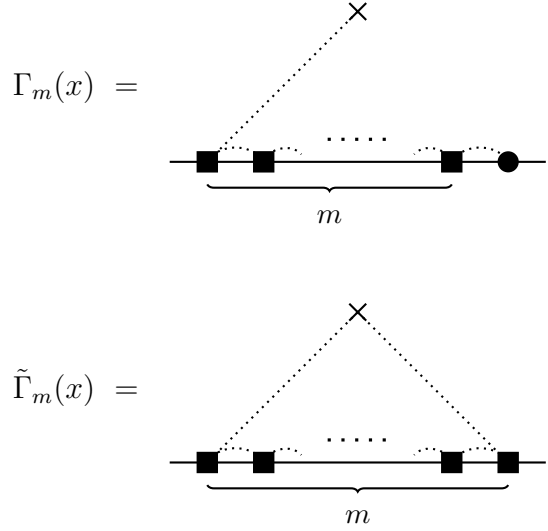
\begin{figure}[h!]
  \centering

    \begin{tikzpicture}[thick]
        \begin{scope}
            \draw (-2.5, 0) -- (2.5, 0);
            \node[cross=4pt] at (0, 2) {};
            \node[rectangle, draw, fill=black] at (-2, 0) {};
            \node[rectangle, draw, fill=black] at (-1.25, 0) {};
            \node[rectangle, draw, fill=black] at (1.25, 0) {};
            \node[dot=8pt] at (2, 0) {};

            \draw[dotted] (0, 2) -- (-2, 0);
            \draw[dotted, bend left=60] (-2, 0) to (-1.25, 0);
            \draw[dotted, bend left=60] (-1.25, 0) to (-0.75, 0.1);
            \draw[dotted, bend right=60] (1.25, 0) to (0.75, 0.1);
            \draw[dotted, bend left=60] (1.25, 0) to (2, 0);
            \draw[loosely dotted, very thick] (-0.4, 0.3) -- (0.4, 0.3);
            \draw [
                thick,
                decoration={
                    brace,
                    mirror,
                    raise=0.5cm
                },
                decorate
            ] (-2, 0.2) -- (1.25, 0.2);
            \node[below] at (-0.375, -0.5) {$m$};
            \node at (-3.75, 1) {\large $\Gamma_m(x)~ =$};
        \end{scope}

        \begin{scope}[yshift=-4cm]
            \draw (-2.5, 0) -- (2.5, 0);
            \node[cross=4pt] at (0, 2) {};
            \node[rectangle, draw, fill=black] at (-2, 0) {};
            \node[rectangle, draw, fill=black] at (-1.25, 0) {};
            \node[rectangle, draw, fill=black] at (1.25, 0) {};
            \node[rectangle, draw, fill=black] at (2, 0) {};

            \draw[dotted] (0, 2) -- (-2, 0);
            \draw[dotted] (0, 2) -- (2, 0);
            \draw[dotted, bend left=60] (-2, 0) to (-1.25, 0);
            \draw[dotted, bend left=60] (-1.25, 0) to (-0.75, 0.1);
            \draw[dotted, bend right=60] (1.25, 0) to (0.75, 0.1);
            \draw[dotted, bend left=60] (1.25, 0) to (2, 0);
            \draw[loosely dotted, very thick] (-0.4, 0.3) -- (0.4, 0.3);
            \draw [
                thick,
                decoration={
                    brace,
                    mirror,
                    raise=0.5cm
                },
                decorate
            ] (-2, 0.2) -- (2, 0.2);
            \node[below] at (0, -0.5) {$m$};
            \node at (-3.75, 1) {\large $\tilde\Gamma_m(x)~ =$};
        \end{scope}
    \end{tikzpicture}
  \caption{The basic diagrams for the evaluation of the one-point functions to all-loop order, with the left one for $\langle \phi(x) \rangle$ and right one for the non-factorizable part of $\langle \phi^2(x) \rangle$.}
\label{fig:onept}
\end{figure}

We consider the defect RG flow in even spacetime dimensions $d = 2n-\eps$ for a single scalar. The action including the two localized deformations $(r=n-1)$ is given as 
\be
    I 
        = 
        \frac{1}{2} \int \d^d x \, (\der \phi)^2  + \frac{h_0}{2} \int_{\mathbb{R}^{2r}} \d^{2r}\, \hat y\, \phi^2 + g_0 \int_{\mathbb{R}^{r}} \d^r \tilde z\, \phi\,,
\ee
where the lower dimensional defect is placed at $x^{r} = x^{r+1}= \cdots = x^{2n-1}=0$ and the higher dimensional defect is placed at $x^{2r} = x^{2r+1}= \cdots = x^{2n-1}=0$. The hatted coordinates cover the higher dimensional defect and the tilde coordinates cover the lower dimensional defect. We work in the minimal subtraction scheme, we can write the bare couplings in terms of the renormalized ones as
\begin{align}
h_0 &=  M^{\eps} \left( h + \frac{\delta h}{\eps} +  \frac{\delta_2 h}{\eps^2}+ \dots \right)\,,\\
g_0 &=  M^{\eps/2} \left( h + \frac{\delta g}{\eps} +  \frac{\delta_2 g}{\eps^2}+ \dots \right)\,.
\end{align}
To obtain the beta functions of the two couplings, it is enough to consider two types of bulk fields, the elementary one $\phi(x)$ and the composite one $\phi^2(x)$. The bare fields and the renormalized ones are related as
\be
\phi (x) = Z_\phi\, [\phi] (x)\,,~~~ \phi^2 (x) = Z_{\phi^2}\,  [\phi^2](x)\,,
\ee
with $Z_\phi$ and $Z_{\phi^2}$ being the wave-function renormalizations, however, for free bulk theory $Z_\phi = Z_{\phi^2} =1$. The physical condition requires that the correlation of the renormalized fields to give finite answer as $\eps \to 0$, in the case of one-point functions, that is
\be\label{eq:finiteness}
\langle\, [\phi](x)\, \rangle = \text{finite}\,,~~~ \langle\, [\phi^2](x) \,\rangle = \text{finite}\,.
\ee
In the current situation, the above correlations can be evaluated to all loop order diagrammatically. Noting that only the completely connected diagrams contribute.\footnote{More precisely, the correlation function in presence of the defects is defined as 
\be
\langle \phi(x) \rangle \equiv \frac{\langle \phi(x) D_L D_H \rangle }{\langle D_L D_H \rangle}\,,~~~
 \langle \phi^2(x) \rangle  \equiv \frac{\langle \phi^2(x) D_L D_H \rangle }{\langle D_L D_H \rangle}\,,
\ee
where $D_{L,H} = e^{-I_{L,H}}$ with $I_{L,H}$ being the action of the lower/higher dimensional deformation.  In this way, it is clear that the partially connected diagrams are excluded.
}
Let us start by considering the first one-point function, for the completely connected ones, it is of the chain type  $\Gamma_m$ as depicted on the left side in figure \ref{fig:onept}. It has $m$ insertions of the higher-dimensional defects and ends on the lower-dimensional defect. 
For simplicity, let us denote the field on the surface defect as $\hat \phi (\hat x)$ and on the line defect  as $\tilde \phi (\tilde x)$.
The chain diagram can be evaluated using the Wick contraction with multiplicity $2^m m!$, 
\begin{align}
    \Gamma_m(x)
        &= 
            (-g_0) \frac{(-h_0/2)^m}{m!}
             \int \d^r\tilde z \int \prod_{i=0}^m \d^{2r} \hat y_i\, \langle \, \phi(x) \prod_{j=0}^{m} \hat \phi^2(\hat y_i)\, \tilde \phi (\tilde z )\, \rangle \no\\
        &=
            -g_0(-h_0)^m\int \frac{\d^{d-r} \tilde p}{(2\pi)^{d-r}}\, f^m_r(|\tilde p_{\parallel,H}|)\, \frac{e^{-\i\, \tilde p x}}{\tilde p^2}
\end{align}
where the function  $f_r$ only depends on the momenta parallel to the higher-dimensional defect while orthogonal to the lower-dimensional one, more explicitly
\be\label{eq:func}
    f_r(|\tilde p_{\parallel,H}|) =(4\pi)^{\frac{\eps-2}{2}}\, |\tilde p_{\parallel,H}| ^{-\eps}\, \Gamma\left(\frac{\eps}{2}\right)\,,
\ee
which is actually independent of the spacetime dimension. 
Summing up all the diagrams gives 
\be \label{eq:onept}
    \langle\, \phi(x)\, \rangle
        = 
        (-g_0) \int \frac{\d^{d-r} \tilde p}{(2\pi)^{d-r} } \, \frac{1} {1+ h_0  f(|\tilde p_{\parallel,H}|) }  \frac{e^{-\i\, \tilde p x}}{\tilde p^2}\,.
\ee

For the evaluation of $ \langle \phi^2(x) \rangle$, there are two types of completely connected diagrams, with either zero or two insertions of the lower-dimensional defect. The diagram with two insertions is essentially a product of two chain-type diagrams coming from the one-point function $\langle \phi(x) \rangle$, summing up all such diagrams gives
\begin{align}
\frac{(-g_0)^2}{2} \sum_{m=0}^\infty \frac{(-h_0/2)^m}{m!} \sum_{i=0}^m 2 {m\choose i} \Gamma_i(x) \Gamma_{m-i}(x) 
= \left( \langle \phi(x) \rangle \right)^2 \,,
\end{align}
where the factor of $2$ in the second sum comes from the pairing of the bulk field and the two insertions of the lower-dimensional defects. The non-factorizable contribution comes from the loop-type diagram  $\tilde \Gamma_m(x)$ depicted on the right side of figure \ref{fig:onept}. They can be evaluated similar to the chain-type diagram with the same multiplicity factor $2^m m!$,
\begin{align}
\tilde \Gamma_m(x) &=\frac{(-h_0/2)^m}{m!} \int \prod_{i=1}^m \d^{2r}\hat y_i \langle \phi^2(x)\prod_{j=1}^m  \hat \phi^2 (\hat y_j) \rangle\no\\
& = (-h_0)^m \int \frac{\d^{d-2r}\hat p}{(2\pi)^{d-2r}} \frac{\d^d p}{(2\pi)^d} \frac{e^{-ix\hat p}}{(\hat p +p)^2 p^2} f^{m-1}_r(|p_{\parallel,H}|) 
\end{align}
where  the function $f_r$ has the same form as in eq. \eqref{eq:func}.
Summing up over all the contributions gives
\begin{align}\label{eq:compositeonept}
\langle\, \phi^2(x)\, \rangle  = (\langle \,\phi(x)\, \rangle )^2 +\int \frac{\d^{d-2r}\hat p}{(2\pi)^{d-2r}}\frac{\d^d p}{(2\pi)^d} \frac{e^{-ix\hat p}}{(\hat p +p)^2 p^2} \frac{ -h_0 }{1+h_0 f_r(|p_{\parallel,H}|)  }\,.
\end{align}
The finiteness condition \eqref{eq:finiteness}  requires that both eq. \eqref{eq:onept}
and \eqref{eq:compositeonept}  are finite on its own, 
\begin{align}
    \frac{M^{\eps/2}}{g_0} + \frac{h_0}{g_0} \frac{M^{-\eps/2}}{2\pi \eps} = \frac{1}{g}\,,~~~ 
    \frac{M^{\eps}}{h_0} +  \frac{1}{2\pi \eps}= \frac{1}{h}
\end{align}
solving them gives
\be
h_0 = \frac{h M^{\epsilon }}{1-\frac{h}{2 \pi  \epsilon }}\,,~~~ g_0 =
   \frac{  g  M^{\epsilon /2} }{1-\frac{h}{2 \pi  \epsilon }}\,.
\ee
Taking the derivatives with respect to the renormalization scale $M$ for the above equations gives the beta functions for the renormalized couplings, one also has to use that fact that the bare coupling is independent of the renormalization scale
\be
\beta_h = \frac{\d h}{\d \ln M} = \frac{h (h-2 \pi  \epsilon )}{2 \pi }\,,
~~~\beta_g = \frac{\d g}{\d \ln M} = \frac{g (h- \pi  \epsilon )}{2 \pi }\,.
\ee
There is still an exact IR fixed point for the higher-dimensional defect, \ie at $h_* = 2\pi \eps$, while there is no IR fixed point for the lower-dimensional defect. 


\section{Computations of anomalous dimensions}\label{app:computaton_anomalous_dim}
\begin{figure}[t]
  \centering
  \begin{tikzpicture}[thick]
      \begin{scope}[xshift=0cm]
          \draw[-] (-2, 0) -- (2, 0);
          \node[cross=4pt] at (-1.75, 0) {};
          \node[cross=4pt, red!80] at (0, 0) {};
          \node[cross=4pt] at (1.75, 0) {};
          \draw[dotted, bend left=60] (-1.75, 0) to (0, 0);
          \draw[dotted, bend left=60] (0, 0) to (1.75, 0);
      \end{scope}

      \begin{scope}[xshift=5cm]
          \draw[-] (-2, 0) -- (2, 0);
          \node[cross=4pt] at (-1.75, 0) {};
          \node[rectangle, draw, fill=black] at (-0.875, 0) {};
          \node[cross=4pt, red!80] at (0, 0) {};
          \node[cross=4pt] at (1.75, 0) {};
          \draw[dotted, bend left=60] (-1.75, 0) to (-0.875, 0);
          \draw[dotted, bend left=45] (-0.875, 0) to (0, 0);
          \draw[dotted, bend right=45] (-0.875, 0) to (0, 0);
          \draw[dotted, bend left=60] (-0.9, 0) to (1.75, 0);
      \end{scope}

      \begin{scope}[xshift=10cm]
          \draw[-] (-2, 0) -- (2, 0);
          \node[cross=4pt] at (-1.75, 0) {};
          \node[dot=8pt] at (-0.875, 0) {};
          \node[cross=4pt, red!80] at (0, 0) {};
          \node[cross=4pt] at (1.75, 0) {};
          \draw[dotted, bend left=60] (-1.75, 0) to (-0.875, 0);
          \draw[dotted, bend left=60] (-0.875, 0) to (0, 0);
          \draw[dotted, bend left=60] (0, 0) to (1.75, 0);
      \end{scope}
  \end{tikzpicture}
  \caption{Feynman diagrams contributing to the three-point function of sub-defect local operators $\langle\, \phi^{I}(\tilde{z}_{1})\,  \phi^{J}(\tilde{z}_{2})\, \phi^{K}\phi^{L}(0)\, \rangle$.
  The dashed line represents the propagator. The black and red cross symbols represent the sub-defect local fundamental field $\phi^{I}$ and composite one $\phi^{K}\phi^{L}$, respectively.
  Also, the circle and square represent the interaction localized on the line and surface defects, respectively.}
\label{fig:anomaloud_dimension_comp}
\end{figure}
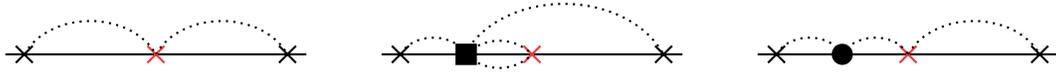

In this appendix, we present the detailed computations of anomalous dimensions of composite operators, which were skipped in the main text. We particularly fix the renormalized matrix $Z_{S}$ which is parameterized by
\begin{align}
    \begin{aligned}
        Z_{S}=\bm{1}_{2\times 2}+
        \begin{pmatrix}
            \delta P & \delta Q \\
            \delta R & \delta S
        \end{pmatrix}
    \end{aligned}
    \ . 
\end{align}
by requiring the renormalization conditions explained around \eqref{eq:three_pt_func}.
For this purpose, we first need to evaluate the following three-point function:
\begin{align}
    \begin{aligned}
        &\langle\, \phi^{I}(\tilde{z}_{1})\,  \phi^{J}(\tilde{z}_{2})\, \phi^{K}\phi^{L}(0)\, \rangle \\
        &\qquad = \langle\, \phi^{I}(\tilde{z}_{1})\,  \phi^{J}(\tilde{z}_{2})\, \phi^{K}\phi^{L}(0)\, \rangle_{0}-\frac{h_{0}}{4!}\int \d^{2}\hat{y}\,\langle\, \phi^{I}(\tilde{z}_{1})\,  \phi^{J}(\tilde{z}_{2})\, \phi^{K}\phi^{L}(0)\, |\vec{\phi}(\hat{y})|^{4}\, \rangle_{0}\\
        &\qquad\qquad\qquad -\frac{g_{0, PQ}}{2}\,\int \d\tilde{z}\,\langle\, \phi^{I}(\tilde{z}_{1})\,  \phi^{J}(\tilde{z}_{2})\, \phi^{K}\phi^{L}(0)\, \phi^{P}\phi^{Q}(\tilde{z})\, \rangle_{0}+(\text{higher order terms})\ . 
    \end{aligned}
\end{align}
(See figure \ref{fig:anomaloud_dimension_comp} for the corresponding Feynman diagrams.)
By using the following integral formulas:
\begin{align}
    \begin{aligned}
        \int\d^{2}\hat{y}\, \frac{1}{|\tilde{z}_{1}-\hat{y}|^{1-\epsilon}|\tilde{z}_{2}-\hat{y}|^{1-\epsilon} |\hat{y}|^{2-2\epsilon}}&=\frac{\pi}{\epsilon}\frac{1}{|\tilde{z}_{1}|\, |\tilde{z}_{2}|}+O(\epsilon^{0})\ ,  \\ 
        \int\d \tilde{z}\frac{1}{|\tilde{z}|^{1-\epsilon} |\tilde{z}-\tilde{z}_{i}|^{1-\epsilon}}&=\frac{4}{\epsilon}\frac{1}{|\tilde{z}_{i}|}+O(\epsilon^{0})\ , 
    \end{aligned}
\end{align}
we can compute the above three-point function as follows:
\begin{align}
    \begin{aligned}
        \langle\, \phi^{I}(\tilde{z}_{1})\,  \phi^{J}(\tilde{z}_{2})\, \phi^{K}\phi^{L}(0)\, \rangle=\frac{C_{\phi\phi}^2}{|\tilde{z}_{1}|\, |\tilde{z}_{2}|}\left(A^{IJKL}-\frac{h_{0}}{48\pi\epsilon}\, B^{IJKL}-\frac{1}{\pi \epsilon}\, C^{IJKL}(g_{0})\right)\ , 
    \end{aligned}
\end{align}
where $A^{IJKL}$, $B^{IJKL}$ and $C^{IJKL}$ are defined by
\begin{align}
    \begin{aligned}
        A^{IJKL}&\equiv\delta^{IK}\delta^{JL}+\delta^{IL}\delta^{JK}\ ,  \\
    B^{IJKL}&\equiv\delta^{IJ}\delta^{KL}+\delta^{IK}\delta^{JL}+\delta^{IL}\delta^{JK}\ ,  \\
        C^{IJKL}(g_{0})&\equiv g_{0, JL}\delta_{IK} +g_{0, JK}\delta_{IL}+g_{0, IL}\delta_{JK} +g_{0, IK}\delta_{JL}\ . 
    \end{aligned}
\end{align}
From this, we can immediately deduce all three-point functions which are necessary to derive anomalous dimensions of composite operators:
\begin{align}
    \begin{aligned} 
        \langle\, \phi^{\alpha}(\tilde{z}_{1})\,  \phi^{\beta}(\tilde{z}_{2})\, \Phi(0)\, \rangle&=\frac{1}{16\pi^{2}\sqrt{m}}\left(2-\frac{m+2}{48\pi\epsilon}h_{0} -\frac{4}{\pi \epsilon}g_{0}\right)\frac{\delta^{\alpha\beta}}{|\tilde{z}_{1}|\, |\tilde{z}_{2}|}+O(\epsilon^{0})\ , \\ 
        \langle\, \phi^{\alpha}(\tilde{z}_{1})\,  \phi^{\beta}(\tilde{z}_{2})\, \Psi(0)\, \rangle&=-\frac{\sqrt{N-m}\, h_{0}}{768\pi^{3} \epsilon}\frac{\delta^{\alpha\beta}}{|\tilde{z}_{1}|\, |\tilde{z}_{2}|}+O(\epsilon^{0})\ , \\ 
        \langle\, \phi^{i}(\tilde{z}_{1})\,  \phi^{j}(\tilde{z}_{2})\, \Phi(0)\, \rangle&=-\frac{\sqrt{m}\, h_{0}}{768\pi^{3} \epsilon}\frac{\delta^{ij}}{|\tilde{z}_{1}|\, |\tilde{z}_{2}|}+O(\epsilon^{0})\ , \\ 
        \langle\, \phi^{i}(\tilde{z}_{1})\,  \phi^{j}(\tilde{z}_{2})\, \Psi(0)\, \rangle&=\frac{1}{16\pi^{2}\sqrt{N-m}}\left(2-\frac{N-m+2}{48\pi\epsilon}h_{0} -\frac{4}{\pi\epsilon}g_{0}'\right)\frac{\delta^{ij}}{|\tilde{z}_{1}|\, |\tilde{z}_{2}|}+O(\epsilon^{0})\ , \\
        \langle\, \phi^{i}(\tilde{z}_{1})\,  \phi^{\alpha}(\tilde{z}_{2})\, \Phi(0)\, \rangle&=
        \langle\, \phi^{i}(\tilde{z}_{1})\,  \phi^{\alpha}(\tilde{z}_{2})\, \Psi(0)\, \rangle=0\  .
    \end{aligned}
\end{align}
Furthermore, by using the results \eqref{eq:wavefunc_ren_fund}, we can obtain the three-point functions consisting of renormalized fields at leading order:
\begin{align}
    \begin{aligned}
        \langle\, [\phi^{\alpha}](\tilde{z}_{1})\,  [\phi^{\beta}](\tilde{z}_{2})\, [\Phi](0)\, \rangle&=\frac{1}{768\sqrt{m}\pi^{3}\epsilon}\left(96\pi\epsilon\, \delta P+(m+2)h+96g \right)+O(\epsilon)\ ,  \\ 
        \langle\, [\phi^{\alpha}](\tilde{z}_{1})\,  [\phi^{\beta}](\tilde{z}_{2})\, [\Psi](0)\, \rangle&=-\frac{1}{768\sqrt{m}\pi^{3}\epsilon}\left(96\pi\epsilon\, \delta R+\sqrt{m(N-m)}h \right)+O(\epsilon) \ , \\
        \langle\, [\phi^{i}](\tilde{z}_{1})\,  [\phi^{j}](\tilde{z}_{2})\, [\Phi](0)\, \rangle&=\frac{1}{768\sqrt{N-m}\pi^{3}\epsilon}\left(96\pi\epsilon\, \delta Q+\sqrt{m(N-m)}h \right)+O(\epsilon) \ , \\
        \langle\, [\phi^{i}](\tilde{z}_{1})\,  [\phi^{j}](\tilde{z}_{2})\, [\Psi](0)\, \rangle&=-\frac{1}{768\sqrt{N-m}\pi^{3}\epsilon}\left(96\pi\epsilon\, \delta S+(N-m+2)h+96g' \right)+O(\epsilon)\ .  
    \end{aligned}
\end{align}
Here, we omit the three-point functions $\langle\, \phi^{i}(\tilde{z}_{1})\,  \phi^{\alpha}(\tilde{z}_{2})\, \Phi(0)\, \rangle$ and $\langle\, \phi^{i}(\tilde{z}_{1})\,  \phi^{\alpha}(\tilde{z}_{2})\, \Psi(0)\, \rangle$ since these both of them do not include divergences. The renormalization conditions, therefore, force the matrix $Z_{S}$ to be fixed as \eqref{eq:renormalization_matrix} in the $\overline{\text{MS}}$ scheme. 
\bibliographystyle{JHEP}
\bibliography{DCFT2}

\providecommand{\href}[2]{#2}\begingroup\raggedright\begin{thebibliography}{100}

\bibitem{Andrei:2018die}
N.~Andrei et~al., \emph{{Boundary and Defect CFT: Open Problems and Applications}}, \href{https://doi.org/10.1088/1751-8121/abb0fe}{\emph{J. Phys. A} {\bfseries 53} (2020) 453002} [\href{https://arxiv.org/abs/1810.05697}{{\ttfamily 1810.05697}}].

\bibitem{Gaiotto:2014kfa}
D.~Gaiotto, A.~Kapustin, N.~Seiberg and B.~Willett, \emph{{Generalized Global Symmetries}}, \href{https://doi.org/10.1007/JHEP02(2015)172}{\emph{JHEP} {\bfseries 02} (2015) 172} [\href{https://arxiv.org/abs/1412.5148}{{\ttfamily 1412.5148}}].

\bibitem{Gaiotto:2017yup}
D.~Gaiotto, A.~Kapustin, Z.~Komargodski and N.~Seiberg, \emph{{Theta, Time Reversal, and Temperature}}, \href{https://doi.org/10.1007/JHEP05(2017)091}{\emph{JHEP} {\bfseries 05} (2017) 091} [\href{https://arxiv.org/abs/1703.00501}{{\ttfamily 1703.00501}}].

\bibitem{Cardy:1984bb}
J.~L. Cardy, \emph{{Conformal Invariance and Surface Critical Behavior}}, \href{https://doi.org/10.1016/0550-3213(84)90241-4}{\emph{Nucl. Phys. B} {\bfseries 240} (1984) 514}.

\bibitem{McAvity:1995zd}
D.~M. McAvity and H.~Osborn, \emph{{Conformal field theories near a boundary in general dimensions}}, \href{https://doi.org/10.1016/0550-3213(95)00476-9}{\emph{Nucl. Phys. B} {\bfseries 455} (1995) 522} [\href{https://arxiv.org/abs/cond-mat/9505127}{{\ttfamily cond-mat/9505127}}].

\bibitem{Kapustin:2005py}
A.~Kapustin, \emph{{Wilson-'t Hooft operators in four-dimensional gauge theories and S-duality}}, \href{https://doi.org/10.1103/PhysRevD.74.025005}{\emph{Phys. Rev. D} {\bfseries 74} (2006) 025005} [\href{https://arxiv.org/abs/hep-th/0501015}{{\ttfamily hep-th/0501015}}].

\bibitem{Liendo:2012hy}
P.~Liendo, L.~Rastelli and B.~C. van Rees, \emph{{The Bootstrap Program for Boundary CFT$_d$}}, \href{https://doi.org/10.1007/JHEP07(2013)113}{\emph{JHEP} {\bfseries 07} (2013) 113} [\href{https://arxiv.org/abs/1210.4258}{{\ttfamily 1210.4258}}].

\bibitem{Billo:2016cpy}
M.~Bill\`o, V.~Gon\c{c}alves, E.~Lauria and M.~Meineri, \emph{{Defects in conformal field theory}}, \href{https://doi.org/10.1007/JHEP04(2016)091}{\emph{JHEP} {\bfseries 04} (2016) 091} [\href{https://arxiv.org/abs/1601.02883}{{\ttfamily 1601.02883}}].

\bibitem{Gadde:2016fbj}
A.~Gadde, \emph{{Conformal constraints on defects}}, \href{https://doi.org/10.1007/JHEP01(2020)038}{\emph{JHEP} {\bfseries 01} (2020) 038} [\href{https://arxiv.org/abs/1602.06354}{{\ttfamily 1602.06354}}].

\bibitem{Lauria:2017wav}
E.~Lauria, M.~Meineri and E.~Trevisani, \emph{{Radial coordinates for defect CFTs}}, \href{https://doi.org/10.1007/JHEP11(2018)148}{\emph{JHEP} {\bfseries 11} (2018) 148} [\href{https://arxiv.org/abs/1712.07668}{{\ttfamily 1712.07668}}].

\bibitem{Lauria:2018klo}
E.~Lauria, M.~Meineri and E.~Trevisani, \emph{{Spinning operators and defects in conformal field theory}}, \href{https://doi.org/10.1007/JHEP08(2019)066}{\emph{JHEP} {\bfseries 08} (2019) 066} [\href{https://arxiv.org/abs/1807.02522}{{\ttfamily 1807.02522}}].

\bibitem{Kobayashi:2018okw}
N.~Kobayashi and T.~Nishioka, \emph{{Spinning conformal defects}}, \href{https://doi.org/10.1007/JHEP09(2018)134}{\emph{JHEP} {\bfseries 09} (2018) 134} [\href{https://arxiv.org/abs/1805.05967}{{\ttfamily 1805.05967}}].

\bibitem{Guha:2018snh}
S.~Guha and B.~Nagaraj, \emph{{Correlators of Mixed Symmetry Operators in Defect CFTs}}, \href{https://doi.org/10.1007/JHEP10(2018)198}{\emph{JHEP} {\bfseries 10} (2018) 198} [\href{https://arxiv.org/abs/1805.12341}{{\ttfamily 1805.12341}}].

\bibitem{Nishioka:2022ook}
T.~Nishioka, Y.~Okuyama and S.~Shimamori, \emph{{Method of images in defect conformal field theories}}, \href{https://doi.org/10.1103/PhysRevD.106.L081701}{\emph{Phys. Rev. D} {\bfseries 106} (2022) L081701} [\href{https://arxiv.org/abs/2205.05370}{{\ttfamily 2205.05370}}].

\bibitem{Kobayashi:2018lil}
N.~Kobayashi, T.~Nishioka, Y.~Sato and K.~Watanabe, \emph{{Towards a $C$-theorem in defect CFT}}, \href{https://doi.org/10.1007/JHEP01(2019)039}{\emph{JHEP} {\bfseries 01} (2019) 039} [\href{https://arxiv.org/abs/1810.06995}{{\ttfamily 1810.06995}}].

\bibitem{Nishioka:2021uef}
T.~Nishioka and Y.~Sato, \emph{{Free energy and defect $C$-theorem in free scalar theory}}, \href{https://doi.org/10.1007/JHEP05(2021)074}{\emph{JHEP} {\bfseries 05} (2021) 074} [\href{https://arxiv.org/abs/2101.02399}{{\ttfamily 2101.02399}}].

\bibitem{Sato:2021eqo}
Y.~Sato, \emph{{Free energy and defect $C$-theorem in free fermion}}, \href{https://doi.org/10.1007/JHEP05(2021)202}{\emph{JHEP} {\bfseries 05} (2021) 202} [\href{https://arxiv.org/abs/2102.11468}{{\ttfamily 2102.11468}}].

\bibitem{Yuan:2022oeo}
M.-K. Yuan and Y.~Zhou, \emph{{Defect localized entropy: Renormalization group and holography}}, \href{https://doi.org/10.1016/j.nuclphysb.2023.116301}{\emph{Nucl. Phys. B} {\bfseries 994} (2023) 116301} [\href{https://arxiv.org/abs/2209.08835}{{\ttfamily 2209.08835}}].

\bibitem{Cuomo:2021rkm}
G.~Cuomo, Z.~Komargodski and A.~Raviv-Moshe, \emph{{Renormalization Group Flows on Line Defects}}, \href{https://doi.org/10.1103/PhysRevLett.128.021603}{\emph{Phys. Rev. Lett.} {\bfseries 128} (2022) 021603} [\href{https://arxiv.org/abs/2108.01117}{{\ttfamily 2108.01117}}].

\bibitem{Casini:2022bsu}
H.~Casini, I.~Salazar~Landea and G.~Torroba, \emph{{Entropic g Theorem in General Spacetime Dimensions}}, \href{https://doi.org/10.1103/PhysRevLett.130.111603}{\emph{Phys. Rev. Lett.} {\bfseries 130} (2023) 111603} [\href{https://arxiv.org/abs/2212.10575}{{\ttfamily 2212.10575}}].

\bibitem{Jensen:2018rxu}
K.~Jensen, A.~O'Bannon, B.~Robinson and R.~Rodgers, \emph{{From the Weyl Anomaly to Entropy of Two-Dimensional Boundaries and Defects}}, \href{https://doi.org/10.1103/PhysRevLett.122.241602}{\emph{Phys. Rev. Lett.} {\bfseries 122} (2019) 241602} [\href{https://arxiv.org/abs/1812.08745}{{\ttfamily 1812.08745}}].

\bibitem{Wang:2021mdq}
Y.~Wang, \emph{{Defect a-theorem and a-maximization}}, \href{https://doi.org/10.1007/JHEP02(2022)061}{\emph{JHEP} {\bfseries 02} (2022) 061} [\href{https://arxiv.org/abs/2101.12648}{{\ttfamily 2101.12648}}].

\bibitem{Casini:2023kyj}
H.~Casini, I.~Salazar~Landea and G.~Torroba, \emph{{Irreversibility, QNEC, and defects}}, \href{https://doi.org/10.1007/JHEP07(2023)004}{\emph{JHEP} {\bfseries 07} (2023) 004} [\href{https://arxiv.org/abs/2303.16935}{{\ttfamily 2303.16935}}].

\bibitem{Harper:2024aku}
J.~Harper, H.~Kanda, T.~Takayanagi and K.~Tasuki, \emph{{The $g$-theorem from Strong Subadditivity}},  \href{https://arxiv.org/abs/2403.19934}{{\ttfamily 2403.19934}}.

\bibitem{Polchinski:1995mt}
J.~Polchinski, \emph{{Dirichlet Branes and Ramond-Ramond charges}}, \href{https://doi.org/10.1103/PhysRevLett.75.4724}{\emph{Phys. Rev. Lett.} {\bfseries 75} (1995) 4724} [\href{https://arxiv.org/abs/hep-th/9510017}{{\ttfamily hep-th/9510017}}].

\bibitem{Recknagel:1997sb}
A.~Recknagel and V.~Schomerus, \emph{{D-branes in Gepner models}}, \href{https://doi.org/10.1016/S0550-3213(98)00468-4}{\emph{Nucl. Phys. B} {\bfseries 531} (1998) 185} [\href{https://arxiv.org/abs/hep-th/9712186}{{\ttfamily hep-th/9712186}}].

\bibitem{Alekseev:1998mc}
A.~Y. Alekseev and V.~Schomerus, \emph{{D-branes in the WZW model}}, \href{https://doi.org/10.1103/PhysRevD.60.061901}{\emph{Phys. Rev. D} {\bfseries 60} (1999) 061901} [\href{https://arxiv.org/abs/hep-th/9812193}{{\ttfamily hep-th/9812193}}].

\bibitem{Recknagel:1998ih}
A.~Recknagel and V.~Schomerus, \emph{{Boundary deformation theory and moduli spaces of D-branes}}, \href{https://doi.org/10.1016/S0550-3213(99)00060-7}{\emph{Nucl. Phys. B} {\bfseries 545} (1999) 233} [\href{https://arxiv.org/abs/hep-th/9811237}{{\ttfamily hep-th/9811237}}].

\bibitem{Schomerus:1999ug}
V.~Schomerus, \emph{{D-branes and deformation quantization}}, \href{https://doi.org/10.1088/1126-6708/1999/06/030}{\emph{JHEP} {\bfseries 06} (1999) 030} [\href{https://arxiv.org/abs/hep-th/9903205}{{\ttfamily hep-th/9903205}}].

\bibitem{Bachas:2001vj}
C.~Bachas, J.~de~Boer, R.~Dijkgraaf and H.~Ooguri, \emph{{Permeable conformal walls and holography}}, \href{https://doi.org/10.1088/1126-6708/2002/06/027}{\emph{JHEP} {\bfseries 06} (2002) 027} [\href{https://arxiv.org/abs/hep-th/0111210}{{\ttfamily hep-th/0111210}}].

\bibitem{DeWolfe:2001pq}
O.~DeWolfe, D.~Z. Freedman and H.~Ooguri, \emph{{Holography and defect conformal field theories}}, \href{https://doi.org/10.1103/PhysRevD.66.025009}{\emph{Phys. Rev. D} {\bfseries 66} (2002) 025009} [\href{https://arxiv.org/abs/hep-th/0111135}{{\ttfamily hep-th/0111135}}].

\bibitem{Karch:2000ct}
A.~Karch and L.~Randall, \emph{{Locally localized gravity}}, \href{https://doi.org/10.1088/1126-6708/2001/05/008}{\emph{JHEP} {\bfseries 05} (2001) 008} [\href{https://arxiv.org/abs/hep-th/0011156}{{\ttfamily hep-th/0011156}}].

\bibitem{Aharony:2003qf}
O.~Aharony, O.~DeWolfe, D.~Z. Freedman and A.~Karch, \emph{{Defect conformal field theory and locally localized gravity}}, \href{https://doi.org/10.1088/1126-6708/2003/07/030}{\emph{JHEP} {\bfseries 07} (2003) 030} [\href{https://arxiv.org/abs/hep-th/0303249}{{\ttfamily hep-th/0303249}}].

\bibitem{Bak:2003jk}
D.~Bak, M.~Gutperle and S.~Hirano, \emph{{A Dilatonic deformation of AdS(5) and its field theory dual}}, \href{https://doi.org/10.1088/1126-6708/2003/05/072}{\emph{JHEP} {\bfseries 05} (2003) 072} [\href{https://arxiv.org/abs/hep-th/0304129}{{\ttfamily hep-th/0304129}}].

\bibitem{DHoker:2007zhm}
E.~D'Hoker, J.~Estes and M.~Gutperle, \emph{{Exact half-BPS Type IIB interface solutions. I. Local solution and supersymmetric Janus}}, \href{https://doi.org/10.1088/1126-6708/2007/06/021}{\emph{JHEP} {\bfseries 06} (2007) 021} [\href{https://arxiv.org/abs/0705.0022}{{\ttfamily 0705.0022}}].

\bibitem{Azeyanagi:2007qj}
T.~Azeyanagi, A.~Karch, T.~Takayanagi and E.~G. Thompson, \emph{{Holographic calculation of boundary entropy}}, \href{https://doi.org/10.1088/1126-6708/2008/03/054}{\emph{JHEP} {\bfseries 03} (2008) 054} [\href{https://arxiv.org/abs/0712.1850}{{\ttfamily 0712.1850}}].

\bibitem{Takayanagi:2011zk}
T.~Takayanagi, \emph{{Holographic Dual of BCFT}}, \href{https://doi.org/10.1103/PhysRevLett.107.101602}{\emph{Phys. Rev. Lett.} {\bfseries 107} (2011) 101602} [\href{https://arxiv.org/abs/1105.5165}{{\ttfamily 1105.5165}}].

\bibitem{Fujita:2011fp}
M.~Fujita, T.~Takayanagi and E.~Tonni, \emph{{Aspects of AdS/BCFT}}, \href{https://doi.org/10.1007/JHEP11(2011)043}{\emph{JHEP} {\bfseries 11} (2011) 043} [\href{https://arxiv.org/abs/1108.5152}{{\ttfamily 1108.5152}}].

\bibitem{Nozaki:2012qd}
M.~Nozaki, T.~Takayanagi and T.~Ugajin, \emph{{Central Charges for BCFTs and Holography}}, \href{https://doi.org/10.1007/JHEP06(2012)066}{\emph{JHEP} {\bfseries 06} (2012) 066} [\href{https://arxiv.org/abs/1205.1573}{{\ttfamily 1205.1573}}].

\bibitem{Bachas:2012bj}
C.~Bachas, I.~Brunner and D.~Roggenkamp, \emph{{A worldsheet extension of O(d,d:Z)}}, \href{https://doi.org/10.1007/JHEP10(2012)039}{\emph{JHEP} {\bfseries 10} (2012) 039} [\href{https://arxiv.org/abs/1205.4647}{{\ttfamily 1205.4647}}].

\bibitem{Jensen:2013lxa}
K.~Jensen and A.~O'Bannon, \emph{{Holography, Entanglement Entropy, and Conformal Field Theories with Boundaries or Defects}}, \href{https://doi.org/10.1103/PhysRevD.88.106006}{\emph{Phys. Rev. D} {\bfseries 88} (2013) 106006} [\href{https://arxiv.org/abs/1309.4523}{{\ttfamily 1309.4523}}].

\bibitem{Bachas:2020yxv}
C.~Bachas, S.~Chapman, D.~Ge and G.~Policastro, \emph{{Energy Reflection and Transmission at 2D Holographic Interfaces}}, \href{https://doi.org/10.1103/PhysRevLett.125.231602}{\emph{Phys. Rev. Lett.} {\bfseries 125} (2020) 231602} [\href{https://arxiv.org/abs/2006.11333}{{\ttfamily 2006.11333}}].

\bibitem{Lemos:2017vnx}
M.~Lemos, P.~Liendo, M.~Meineri and S.~Sarkar, \emph{{Universality at large transverse spin in defect CFT}}, \href{https://doi.org/10.1007/JHEP09(2018)091}{\emph{JHEP} {\bfseries 09} (2018) 091} [\href{https://arxiv.org/abs/1712.08185}{{\ttfamily 1712.08185}}].

\bibitem{Liendo:2019jpu}
P.~Liendo, Y.~Linke and V.~Schomerus, \emph{{A Lorentzian inversion formula for defect CFT}}, \href{https://doi.org/10.1007/JHEP08(2020)163}{\emph{JHEP} {\bfseries 08} (2020) 163} [\href{https://arxiv.org/abs/1903.05222}{{\ttfamily 1903.05222}}].

\bibitem{Ghosh:2021ruh}
K.~Ghosh, A.~Kaviraj and M.~F. Paulos, \emph{{Charging up the functional bootstrap}}, \href{https://doi.org/10.1007/JHEP10(2021)116}{\emph{JHEP} {\bfseries 10} (2021) 116} [\href{https://arxiv.org/abs/2107.00041}{{\ttfamily 2107.00041}}].

\bibitem{Yamaguchi:2016pbj}
S.~Yamaguchi, \emph{{The \ensuremath{\epsilon}-expansion of the codimension two twist defect from conformal field theory}}, \href{https://doi.org/10.1093/ptep/ptw115}{\emph{PTEP} {\bfseries 2016} (2016) 091B01} [\href{https://arxiv.org/abs/1607.05551}{{\ttfamily 1607.05551}}].

\bibitem{Nishioka:2022odm}
T.~Nishioka, Y.~Okuyama and S.~Shimamori, \emph{{Comments on epsilon expansion of the O(N) model with boundary}}, \href{https://doi.org/10.1007/JHEP03(2023)051}{\emph{JHEP} {\bfseries 03} (2023) 051} [\href{https://arxiv.org/abs/2212.04078}{{\ttfamily 2212.04078}}].

\bibitem{Nishioka:2022qmj}
T.~Nishioka, Y.~Okuyama and S.~Shimamori, \emph{{The epsilon expansion of the O(N) model with line defect from conformal field theory}}, \href{https://doi.org/10.1007/JHEP03(2023)203}{\emph{JHEP} {\bfseries 03} (2023) 203} [\href{https://arxiv.org/abs/2212.04076}{{\ttfamily 2212.04076}}].

\bibitem{Gaiotto:2013nva}
D.~Gaiotto, D.~Mazac and M.~F. Paulos, \emph{{Bootstrapping the 3d Ising twist defect}}, \href{https://doi.org/10.1007/JHEP03(2014)100}{\emph{JHEP} {\bfseries 03} (2014) 100} [\href{https://arxiv.org/abs/1310.5078}{{\ttfamily 1310.5078}}].

\bibitem{Bianchi:2021snj}
L.~Bianchi, A.~Chalabi, V.~Proch\'azka, B.~Robinson and J.~Sisti, \emph{{Monodromy defects in free field theories}}, \href{https://doi.org/10.1007/JHEP08(2021)013}{\emph{JHEP} {\bfseries 08} (2021) 013} [\href{https://arxiv.org/abs/2104.01220}{{\ttfamily 2104.01220}}].

\bibitem{Gimenez-Grau:2021wiv}
A.~Gimenez-Grau and P.~Liendo, \emph{{Bootstrapping monodromy defects in the Wess-Zumino model}}, \href{https://doi.org/10.1007/JHEP05(2022)185}{\emph{JHEP} {\bfseries 05} (2022) 185} [\href{https://arxiv.org/abs/2108.05107}{{\ttfamily 2108.05107}}].

\bibitem{Giombi:2021uae}
S.~Giombi, E.~Helfenberger, Z.~Ji and H.~Khanchandani, \emph{{Monodromy defects from hyperbolic space}}, \href{https://doi.org/10.1007/JHEP02(2022)041}{\emph{JHEP} {\bfseries 02} (2022) 041} [\href{https://arxiv.org/abs/2102.11815}{{\ttfamily 2102.11815}}].

\bibitem{Collier:2021ngi}
S.~Collier, D.~Mazac and Y.~Wang, \emph{{Bootstrapping boundaries and branes}}, \href{https://doi.org/10.1007/JHEP02(2023)019}{\emph{JHEP} {\bfseries 02} (2023) 019} [\href{https://arxiv.org/abs/2112.00750}{{\ttfamily 2112.00750}}].

\bibitem{Soderberg:2021kne}
A.~S\"oderberg, \emph{{Fusion of conformal defects in four dimensions}}, \href{https://doi.org/10.1007/JHEP04(2021)087}{\emph{JHEP} {\bfseries 04} (2021) 087} [\href{https://arxiv.org/abs/2102.00718}{{\ttfamily 2102.00718}}].

\bibitem{Bissi:2022bgu}
A.~Bissi, P.~Dey, J.~Sisti and A.~S\"oderberg, \emph{{Interacting conformal scalar in a wedge}}, \href{https://doi.org/10.1007/JHEP10(2022)060}{\emph{JHEP} {\bfseries 10} (2022) 060} [\href{https://arxiv.org/abs/2206.06326}{{\ttfamily 2206.06326}}].

\bibitem{SoderbergRousu:2023zyj}
A.~S\"oderberg~Rousu, \emph{{Fusion of conformal defects in interacting theories}}, \href{https://doi.org/10.1007/JHEP10(2023)183}{\emph{JHEP} {\bfseries 10} (2023) 183} [\href{https://arxiv.org/abs/2304.10239}{{\ttfamily 2304.10239}}].

\bibitem{Popov:2022nfq}
F.~K. Popov and Y.~Wang, \emph{{Non-perturbative defects in tensor models from melonic trees}}, \href{https://doi.org/10.1007/JHEP11(2022)057}{\emph{JHEP} {\bfseries 11} (2022) 057} [\href{https://arxiv.org/abs/2206.14206}{{\ttfamily 2206.14206}}].

\bibitem{Brax:2023goj}
P.~Brax and S.~Fichet, \emph{{Casimir Forces in CFT with Defects and Boundaries}},  \href{https://arxiv.org/abs/2312.02281}{{\ttfamily 2312.02281}}.

\bibitem{McAvity:1993ue}
D.~M. McAvity and H.~Osborn, \emph{{Energy momentum tensor in conformal field theories near a boundary}}, \href{https://doi.org/10.1016/0550-3213(93)90005-A}{\emph{Nucl. Phys. B} {\bfseries 406} (1993) 655} [\href{https://arxiv.org/abs/hep-th/9302068}{{\ttfamily hep-th/9302068}}].

\bibitem{Herzog:2019bom}
C.~P. Herzog and I.~Shamir, \emph{{On Marginal Operators in Boundary Conformal Field Theory}}, \href{https://doi.org/10.1007/JHEP10(2019)088}{\emph{JHEP} {\bfseries 10} (2019) 088} [\href{https://arxiv.org/abs/1906.11281}{{\ttfamily 1906.11281}}].

\bibitem{Herzog:2020lel}
C.~P. Herzog and N.~Kobayashi, \emph{{The $O(N)$ model with $\phi^6$ potential in ${\mathbb R}^2 \times {\mathbb R}^+$}}, \href{https://doi.org/10.1007/JHEP09(2020)126}{\emph{JHEP} {\bfseries 09} (2020) 126} [\href{https://arxiv.org/abs/2005.07863}{{\ttfamily 2005.07863}}].

\bibitem{Herzog:2022jlx}
C.~P. Herzog and V.~Schaub, \emph{{Fermions in boundary conformal field theory: crossing symmetry and E-expansion}}, \href{https://doi.org/10.1007/JHEP02(2023)129}{\emph{JHEP} {\bfseries 02} (2023) 129} [\href{https://arxiv.org/abs/2209.05511}{{\ttfamily 2209.05511}}].

\bibitem{Herzog:2023dop}
C.~P. Herzog and V.~Schaub, \emph{{Tilting space of boundary conformal field theories}}, \href{https://doi.org/10.1103/PhysRevD.109.L061701}{\emph{Phys. Rev. D} {\bfseries 109} (2024) L061701} [\href{https://arxiv.org/abs/2301.10789}{{\ttfamily 2301.10789}}].

\bibitem{Bartlett-Tisdall:2023ghh}
S.~Bartlett-Tisdall, C.~P. Herzog and V.~Schaub, \emph{{Bootstrapping boundary QED. Part I}}, \href{https://doi.org/10.1007/JHEP05(2024)235}{\emph{JHEP} {\bfseries 05} (2024) 235} [\href{https://arxiv.org/abs/2312.07692}{{\ttfamily 2312.07692}}].

\bibitem{Bissi:2018mcq}
A.~Bissi, T.~Hansen and A.~S\"oderberg, \emph{{Analytic Bootstrap for Boundary CFT}}, \href{https://doi.org/10.1007/JHEP01(2019)010}{\emph{JHEP} {\bfseries 01} (2019) 010} [\href{https://arxiv.org/abs/1808.08155}{{\ttfamily 1808.08155}}].

\bibitem{Prochazka:2019fah}
V.~Proch\'azka and A.~S\"oderberg, \emph{{Composite operators near the boundary}}, \href{https://doi.org/10.1007/JHEP03(2020)114}{\emph{JHEP} {\bfseries 03} (2020) 114} [\href{https://arxiv.org/abs/1912.07505}{{\ttfamily 1912.07505}}].

\bibitem{Prochazka:2020vog}
V.~Proch\'azka and A.~S\"oderberg, \emph{{Spontaneous symmetry breaking in free theories with boundary potentials}},  \href{https://arxiv.org/abs/2012.00701}{{\ttfamily 2012.00701}}.

\bibitem{DiPietro:2019hqe}
L.~Di~Pietro, D.~Gaiotto, E.~Lauria and J.~Wu, \emph{{3d Abelian Gauge Theories at the Boundary}}, \href{https://doi.org/10.1007/JHEP05(2019)091}{\emph{JHEP} {\bfseries 05} (2019) 091} [\href{https://arxiv.org/abs/1902.09567}{{\ttfamily 1902.09567}}].

\bibitem{Behan:2020nsf}
C.~Behan, L.~Di~Pietro, E.~Lauria and B.~C. Van~Rees, \emph{{Bootstrapping boundary-localized interactions}}, \href{https://doi.org/10.1007/JHEP12(2020)182}{\emph{JHEP} {\bfseries 12} (2020) 182} [\href{https://arxiv.org/abs/2009.03336}{{\ttfamily 2009.03336}}].

\bibitem{DiPietro:2020fya}
L.~Di~Pietro, E.~Lauria and P.~Niro, \emph{{3d large $N$ vector models at the boundary}}, \href{https://doi.org/10.21468/SciPostPhys.11.3.050}{\emph{SciPost Phys.} {\bfseries 11} (2021) 050} [\href{https://arxiv.org/abs/2012.07733}{{\ttfamily 2012.07733}}].

\bibitem{Behan:2021tcn}
C.~Behan, L.~Di~Pietro, E.~Lauria and B.~C. van Rees, \emph{{Bootstrapping boundary-localized interactions II. Minimal models at the boundary}}, \href{https://doi.org/10.1007/JHEP03(2022)146}{\emph{JHEP} {\bfseries 03} (2022) 146} [\href{https://arxiv.org/abs/2111.04747}{{\ttfamily 2111.04747}}].

\bibitem{DiPietro:2023gzi}
L.~Di~Pietro, E.~Lauria and P.~Niro, \emph{{Conformal boundary conditions for a 4d scalar field}}, \href{https://doi.org/10.21468/SciPostPhys.16.4.090}{\emph{SciPost Phys.} {\bfseries 16} (2024) 090} [\href{https://arxiv.org/abs/2312.11633}{{\ttfamily 2312.11633}}].

\bibitem{Giombi:2019enr}
S.~Giombi and H.~Khanchandani, \emph{{$O(N)$ models with boundary interactions and their long range generalizations}}, \href{https://doi.org/10.1007/JHEP08(2020)010}{\emph{JHEP} {\bfseries 08} (2020) 010} [\href{https://arxiv.org/abs/1912.08169}{{\ttfamily 1912.08169}}].

\bibitem{Cuomo:2021cnb}
G.~Cuomo, M.~Mezei and A.~Raviv-Moshe, \emph{{Boundary conformal field theory at large charge}}, \href{https://doi.org/10.1007/JHEP10(2021)143}{\emph{JHEP} {\bfseries 10} (2021) 143} [\href{https://arxiv.org/abs/2108.06579}{{\ttfamily 2108.06579}}].

\bibitem{Harribey:2023xyv}
S.~Harribey, I.~R. Klebanov and Z.~Sun, \emph{{Boundaries and interfaces with localized cubic interactions in the O(N) model}}, \href{https://doi.org/10.1007/JHEP10(2023)017}{\emph{JHEP} {\bfseries 10} (2023) 017} [\href{https://arxiv.org/abs/2307.00072}{{\ttfamily 2307.00072}}].

\bibitem{Metlitski:2020cqy}
M.~A. Metlitski, \emph{{Boundary criticality of the O(N) model in d = 3 critically revisited}}, \href{https://doi.org/10.21468/SciPostPhys.12.4.131}{\emph{SciPost Phys.} {\bfseries 12} (2022) 131} [\href{https://arxiv.org/abs/2009.05119}{{\ttfamily 2009.05119}}].

\bibitem{Toldin:2021kun}
F.~P. Toldin and M.~A. Metlitski, \emph{{Boundary Criticality of the 3D O(N) Model: From Normal to Extraordinary}}, \href{https://doi.org/10.1103/PhysRevLett.128.215701}{\emph{Phys. Rev. Lett.} {\bfseries 128} (2022) 215701} [\href{https://arxiv.org/abs/2111.03613}{{\ttfamily 2111.03613}}].

\bibitem{Shen:2024ixw}
X.~Shen, Z.~Wu and S.-K. Jian, \emph{{New boundary criticality in topological phases}},  \href{https://arxiv.org/abs/2407.15916}{{\ttfamily 2407.15916}}.

\bibitem{Billo:2013jda}
M.~Bill\'o, M.~Caselle, D.~Gaiotto, F.~Gliozzi, M.~Meineri and R.~Pellegrini, \emph{{Line defects in the 3d Ising model}}, \href{https://doi.org/10.1007/JHEP07(2013)055}{\emph{JHEP} {\bfseries 07} (2013) 055} [\href{https://arxiv.org/abs/1304.4110}{{\ttfamily 1304.4110}}].

\bibitem{Soderberg:2017oaa}
A.~S\"oderberg, \emph{{Anomalous Dimensions in the WF O($N$) Model with a Monodromy Line Defect}}, \href{https://doi.org/10.1007/JHEP03(2018)058}{\emph{JHEP} {\bfseries 03} (2018) 058} [\href{https://arxiv.org/abs/1706.02414}{{\ttfamily 1706.02414}}].

\bibitem{Lauria:2020emq}
E.~Lauria, P.~Liendo, B.~C. Van~Rees and X.~Zhao, \emph{{Line and surface defects for the free scalar field}}, \href{https://doi.org/10.1007/JHEP01(2021)060}{\emph{JHEP} {\bfseries 01} (2021) 060} [\href{https://arxiv.org/abs/2005.02413}{{\ttfamily 2005.02413}}].

\bibitem{Cuomo:2021kfm}
G.~Cuomo, Z.~Komargodski and M.~Mezei, \emph{{Localized magnetic field in the O(N) model}}, \href{https://doi.org/10.1007/JHEP02(2022)134}{\emph{JHEP} {\bfseries 02} (2022) 134} [\href{https://arxiv.org/abs/2112.10634}{{\ttfamily 2112.10634}}].

\bibitem{Cuomo:2022xgw}
G.~Cuomo, Z.~Komargodski, M.~Mezei and A.~Raviv-Moshe, \emph{{Spin impurities, Wilson lines and semiclassics}}, \href{https://doi.org/10.1007/JHEP06(2022)112}{\emph{JHEP} {\bfseries 06} (2022) 112} [\href{https://arxiv.org/abs/2202.00040}{{\ttfamily 2202.00040}}].

\bibitem{Giombi:2022vnz}
S.~Giombi, E.~Helfenberger and H.~Khanchandani, \emph{{Line defects in fermionic CFTs}}, \href{https://doi.org/10.1007/JHEP08(2023)224}{\emph{JHEP} {\bfseries 08} (2023) 224} [\href{https://arxiv.org/abs/2211.11073}{{\ttfamily 2211.11073}}].

\bibitem{Gimenez-Grau:2022czc}
A.~Gimenez-Grau, E.~Lauria, P.~Liendo and P.~van Vliet, \emph{{Bootstrapping line defects with O(2) global symmetry}}, \href{https://doi.org/10.1007/JHEP11(2022)018}{\emph{JHEP} {\bfseries 11} (2022) 018} [\href{https://arxiv.org/abs/2208.11715}{{\ttfamily 2208.11715}}].

\bibitem{Aharony:2022ntz}
O.~Aharony, G.~Cuomo, Z.~Komargodski, M.~Mezei and A.~Raviv-Moshe, \emph{{Phases of Wilson Lines in Conformal Field Theories}}, \href{https://doi.org/10.1103/PhysRevLett.130.151601}{\emph{Phys. Rev. Lett.} {\bfseries 130} (2023) 151601} [\href{https://arxiv.org/abs/2211.11775}{{\ttfamily 2211.11775}}].

\bibitem{Aharony:2023amq}
O.~Aharony, G.~Cuomo, Z.~Komargodski, M.~Mezei and A.~Raviv-Moshe, \emph{{Phases of Wilson lines: conformality and screening}}, \href{https://doi.org/10.1007/JHEP12(2023)183}{\emph{JHEP} {\bfseries 12} (2023) 183} [\href{https://arxiv.org/abs/2310.00045}{{\ttfamily 2310.00045}}].

\bibitem{Dey:2024ilw}
P.~Dey and K.~Ghosh, \emph{{Bootstrapping conformal defect operators on a line}},  \href{https://arxiv.org/abs/2404.06576}{{\ttfamily 2404.06576}}.

\bibitem{Herzog:2022jqv}
C.~P. Herzog and A.~Shrestha, \emph{{Conformal surface defects in Maxwell theory are trivial}}, \href{https://doi.org/10.1007/JHEP08(2022)282}{\emph{JHEP} {\bfseries 08} (2022) 282} [\href{https://arxiv.org/abs/2202.09180}{{\ttfamily 2202.09180}}].

\bibitem{Wang:2020xkc}
Y.~Wang, \emph{{Surface defect, anomalies and b-extremization}}, \href{https://doi.org/10.1007/JHEP11(2021)122}{\emph{JHEP} {\bfseries 11} (2021) 122} [\href{https://arxiv.org/abs/2012.06574}{{\ttfamily 2012.06574}}].

\bibitem{Cuomo:2023qvp}
G.~Cuomo and S.~Zhang, \emph{{Spontaneous symmetry breaking on surface defects}}, \href{https://doi.org/10.1007/JHEP03(2024)022}{\emph{JHEP} {\bfseries 03} (2024) 022} [\href{https://arxiv.org/abs/2306.00085}{{\ttfamily 2306.00085}}].

\bibitem{Shachar:2022fqk}
T.~Shachar, R.~Sinha and M.~Smolkin, \emph{{RG flows on two-dimensional spherical defects}}, \href{https://doi.org/10.21468/SciPostPhys.15.6.240}{\emph{SciPost Phys.} {\bfseries 15} (2023) 240} [\href{https://arxiv.org/abs/2212.08081}{{\ttfamily 2212.08081}}].

\bibitem{Giombi:2023dqs}
S.~Giombi and B.~Liu, \emph{{Notes on a surface defect in the O(N) model}}, \href{https://doi.org/10.1007/JHEP12(2023)004}{\emph{JHEP} {\bfseries 12} (2023) 004} [\href{https://arxiv.org/abs/2305.11402}{{\ttfamily 2305.11402}}].

\bibitem{Raviv-Moshe:2023yvq}
A.~Raviv-Moshe and S.~Zhong, \emph{{Phases of surface defects in Scalar Field Theories}}, \href{https://doi.org/10.1007/JHEP08(2023)143}{\emph{JHEP} {\bfseries 08} (2023) 143} [\href{https://arxiv.org/abs/2305.11370}{{\ttfamily 2305.11370}}].

\bibitem{Trepanier:2023tvb}
M.~Tr\'epanier, \emph{{Surface defects in the O(N) model}}, \href{https://doi.org/10.1007/JHEP09(2023)074}{\emph{JHEP} {\bfseries 09} (2023) 074} [\href{https://arxiv.org/abs/2305.10486}{{\ttfamily 2305.10486}}].

\bibitem{Krishnan:2023cff}
A.~Krishnan and M.~A. Metlitski, \emph{{A plane defect in the 3d O(N) model}}, \href{https://doi.org/10.21468/SciPostPhys.15.3.090}{\emph{SciPost Phys.} {\bfseries 15} (2023) 090} [\href{https://arxiv.org/abs/2301.05728}{{\ttfamily 2301.05728}}].

\bibitem{hagiwara2016surface}
K.~Hagiwara, Y.~Ohtsubo, M.~Matsunami, S.-i. Ideta, K.~Tanaka, H.~Miyazaki et~al., \emph{Surface kondo effect and non-trivial metallic state of the kondo insulator ybb12}, {\emph{Nature communications} {\bfseries 7} (2016) 12690}.

\bibitem{Cardy_1983}
J.~L. Cardy, \emph{Critical behaviour at an edge}, \href{https://doi.org/10.1088/0305-4470/16/15/026}{\emph{Journal of Physics A: Mathematical and General} {\bfseries 16} (1983) 3617}.

\bibitem{Antunes:2021qpy}
A.~Antunes, \emph{{Conformal bootstrap near the edge}}, \href{https://doi.org/10.1007/JHEP10(2021)057}{\emph{JHEP} {\bfseries 10} (2021) 057} [\href{https://arxiv.org/abs/2103.03132}{{\ttfamily 2103.03132}}].

\bibitem{Shimamori:2024yms}
S.~Shimamori, \emph{{Conformal field theory with composite defect}},  \href{https://arxiv.org/abs/2404.08411}{{\ttfamily 2404.08411}}.

\bibitem{Eisenriegler:1988}
E.~Eisenriegler and H.~W. Diehl, \emph{Surface critical behavior of tricritical systems}, \href{https://doi.org/10.1103/PhysRevB.37.5257}{\emph{Phys. Rev. B} {\bfseries 37} (1988) 5257}.

\bibitem{Jepsen:2020czw}
C.~B. Jepsen, I.~R. Klebanov and F.~K. Popov, \emph{{RG limit cycles and unconventional fixed points in perturbative QFT}}, \href{https://doi.org/10.1103/PhysRevD.103.046015}{\emph{Phys. Rev. D} {\bfseries 103} (2021) 046015} [\href{https://arxiv.org/abs/2010.15133}{{\ttfamily 2010.15133}}].

\bibitem{Barnes:2004jj}
E.~Barnes, K.~A. Intriligator, B.~Wecht and J.~Wright, \emph{{Evidence for the strongest version of the 4d a-theorem, via a-maximization along RG flows}}, \href{https://doi.org/10.1016/j.nuclphysb.2004.09.016}{\emph{Nucl. Phys. B} {\bfseries 702} (2004) 131} [\href{https://arxiv.org/abs/hep-th/0408156}{{\ttfamily hep-th/0408156}}].

\bibitem{Nishioka:2018khk}
T.~Nishioka, \emph{{Entanglement entropy: holography and renormalization group}}, \href{https://doi.org/10.1103/RevModPhys.90.035007}{\emph{Rev. Mod. Phys.} {\bfseries 90} (2018) 035007} [\href{https://arxiv.org/abs/1801.10352}{{\ttfamily 1801.10352}}].

\bibitem{Cardy:1988cwa}
J.~L. Cardy, \emph{{Is There a c Theorem in Four-Dimensions?}}, \href{https://doi.org/10.1016/0370-2693(88)90054-8}{\emph{Phys. Lett. B} {\bfseries 215} (1988) 749}.

\bibitem{Klebanov:2011gs}
I.~R. Klebanov, S.~S. Pufu and B.~R. Safdi, \emph{{F-Theorem without Supersymmetry}}, \href{https://doi.org/10.1007/JHEP10(2011)038}{\emph{JHEP} {\bfseries 10} (2011) 038} [\href{https://arxiv.org/abs/1105.4598}{{\ttfamily 1105.4598}}].

\bibitem{Gaiotto:2014gha}
D.~Gaiotto, \emph{{Boundary F-maximization}},  \href{https://arxiv.org/abs/1403.8052}{{\ttfamily 1403.8052}}.

\bibitem{Gaberdiel:2008fn}
M.~R. Gaberdiel, A.~Konechny and C.~Schmidt-Colinet, \emph{{Conformal perturbation theory beyond the leading order}}, \href{https://doi.org/10.1088/1751-8113/42/10/105402}{\emph{J. Phys. A} {\bfseries 42} (2009) 105402} [\href{https://arxiv.org/abs/0811.3149}{{\ttfamily 0811.3149}}].

\bibitem{Bosschaert:2021ooy}
M.~M. Bosschaert, C.~B. Jepsen and F.~K. Popov, \emph{{Chaotic RG flow in tensor models}}, \href{https://doi.org/10.1103/PhysRevD.105.065021}{\emph{Phys. Rev. D} {\bfseries 105} (2022) 065021} [\href{https://arxiv.org/abs/2112.09088}{{\ttfamily 2112.09088}}].

\bibitem{Friedan:2003yc}
D.~Friedan and A.~Konechny, \emph{{On the boundary entropy of one-dimensional quantum systems at low temperature}}, \href{https://doi.org/10.1103/PhysRevLett.93.030402}{\emph{Phys. Rev. Lett.} {\bfseries 93} (2004) 030402} [\href{https://arxiv.org/abs/hep-th/0312197}{{\ttfamily hep-th/0312197}}].

\bibitem{Jensen:2015swa}
K.~Jensen and A.~O'Bannon, \emph{{Constraint on Defect and Boundary Renormalization Group Flows}}, \href{https://doi.org/10.1103/PhysRevLett.116.091601}{\emph{Phys. Rev. Lett.} {\bfseries 116} (2016) 091601} [\href{https://arxiv.org/abs/1509.02160}{{\ttfamily 1509.02160}}].

\bibitem{Komargodski:2011vj}
Z.~Komargodski and A.~Schwimmer, \emph{{On Renormalization Group Flows in Four Dimensions}}, \href{https://doi.org/10.1007/JHEP12(2011)099}{\emph{JHEP} {\bfseries 12} (2011) 099} [\href{https://arxiv.org/abs/1107.3987}{{\ttfamily 1107.3987}}].

\bibitem{Komargodski:2011xv}
Z.~Komargodski, \emph{{The Constraints of Conformal Symmetry on RG Flows}}, \href{https://doi.org/10.1007/JHEP07(2012)069}{\emph{JHEP} {\bfseries 07} (2012) 069} [\href{https://arxiv.org/abs/1112.4538}{{\ttfamily 1112.4538}}].

\bibitem{Harribey:2024gjn}
S.~Harribey, W.~H. Pannell and A.~Stergiou, \emph{{Multiscalar Critical Models with Localised Cubic Interactions}},  \href{https://arxiv.org/abs/2407.20326}{{\ttfamily 2407.20326}}.

\end{thebibliography}\endgroup

\end{document}